\documentclass{article}

\usepackage[english]{babel}

\usepackage[letterpaper,top=2cm,bottom=2cm,left=3cm,right=3cm,marginparwidth=1.75cm]{geometry}

\usepackage{amsmath}
\usepackage{amssymb}
\usepackage{graphicx}
\usepackage[colorlinks=true, allcolors=blue]{hyperref}
\usepackage[square,sort,comma,numbers]{natbib}

\usepackage{tikz}
\usepackage{amsmath}
\usepackage{graphicx}
\usepackage{fullpage}
\usepackage[colorlinks=true,allcolors=blue]{hyperref}
\usepackage[capitalise]{cleveref}
\usepackage[utf8]{inputenc}
\usepackage{siunitx}
\usepackage{url}
\usepackage{rotating}
\usepackage{epsfig,graphics,rotate}
\usepackage{wrapfig}
\usepackage[caption=false]{subfig}
\usepackage{bm}
\usepackage{amssymb}
\usepackage{amsmath}
\usepackage{amsfonts}
\usepackage{booktabs}
\usepackage{microtype}
\usepackage{multirow, makecell}
\usepackage{lipsum}
\usepackage{xspace}
\usepackage{comment}
\usepackage{floatrow}
\usepackage{ragged2e}
\usepackage{nicefrac, xfrac}
\usepackage{mathtools}
\usepackage{relsize} 
\usepackage{float}
\usepackage{psvectorian}
\usepackage{bbding}

\title{IceCube Results and Perspective for Neutrinos from LHAASO Sources}
\author{Ke Fang and Francis Halzen}

\begin{document}
\maketitle

\begin{abstract}
We briefly review the main results of the IceCube Neutrino Observatory one decade after the discovery of cosmic neutrinos. We emphasize the importance of multimessenger observations, most prominently for the discovery of neutrinos from our own Galaxy. We model the flux from the Galactic plane produced by Galactic cosmic rays interacting with the interstellar medium and discuss the perspectives of understanding the TeV-PeV emission of the Galactic plane by combining neutrino and gamma-ray observations.
We draw attention to the interesting fact that the neutrino flux from the Galaxy is not a dominant feature of the neutrino sky, unlike the case in any other wavelength of light. Finally, we review the attempts to identify PeVatrons by confronting the neutrino and gamma-ray emission of Galactic sources, including those observed by LHAASO. We end with a discussion of searches for neutrinos from LHAASO's extragalactic transient source gamma-ray burst 221009A.  
\end{abstract}

\section{IceCube: The First Decade of High-Energy Neutrino Astronomy}

Conceptualized more than half a century ago, high-energy neutrino astronomy became a reality when the IceCube project transformed a cubic kilometer of transparent natural Antarctic ice one mile below the geographic South Pole into the largest particle detector ever built~\cite{IceCube:2016zyt}.
IceCube discovered neutrinos in the $\rm TeV \sim PeV$ energy range originating beyond our galaxy, opening a new window for astronomy~\cite{Aartsen:2013jdh,IceCube:2014stg}.
The observed energy density of neutrinos in the extreme universe exceeds the energy in gamma rays~\cite{Fang:2022trf}; it even outshines the neutrino flux from the nearby sources in our own galaxy~\cite{IceCube:2023ame}.
The Galactic plane only appears as a faint glow at a level of ten percent of the extragalactic flux, in sharp contrast with what is found for any other wavelength of light, where the Milky Way is the dominant feature in the sky.
This observation implies the existence of sources of high-energy neutrinos in other galaxies that are not present in our own~\cite{Fang:2023azx}.

After accumulating a decade of data with a detector with gradually improved sensitivity, the first high-energy neutrino sources emerged in the neutrino sky: the active galaxies NGC 1068, NGC 4151, PKS 1424+240~\cite{IceCube:2022der,IceCube:2023jds}, and TXS 0506+056~\cite{IceCube:2018dnn,IceCube:2018cha}.
The observations point to the acceleration of protons and the production of neutrinos in the obscured dense cores surrounding the supermassive black holes at their center, typically within a distance of only $10 \sim 100$ Schwarzschild radii~\cite{IceCube:2022der}.
TXS 0506+056 had previously been identified as a neutrino source in a multimessenger campaign triggered by a neutrino of 290~TeV energy, IC170922, and by the independent observation of a neutrino burst from the same source in archival IceCube data in 2014~\cite{IceCube:2018dnn,IceCube:2018cha}.

\begin{figure}[ht]
\centering
\includegraphics[width=0.6\linewidth, trim=0 150 0 150, clip=true]{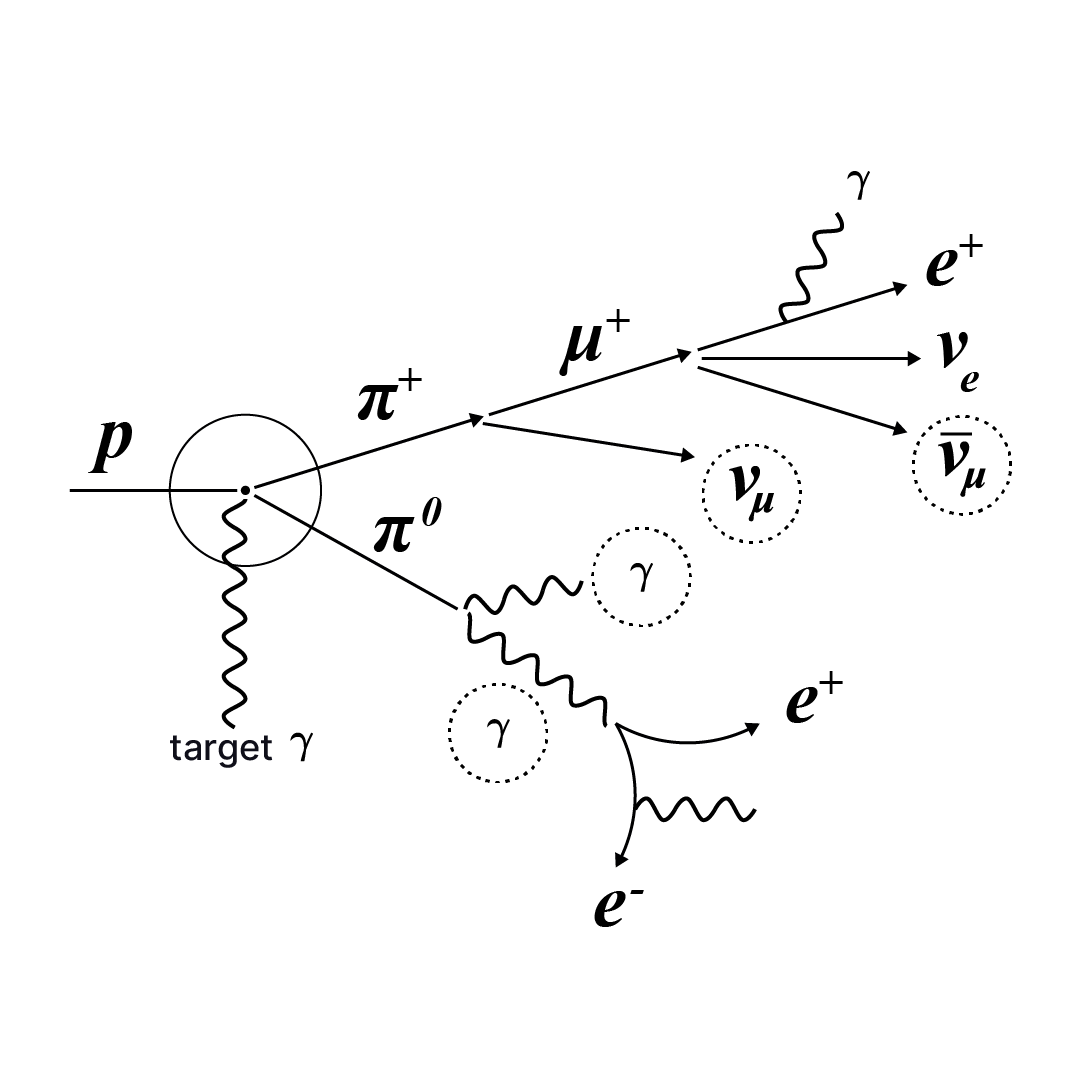}
\protect\caption{\textbf{Flow diagram showing the production of charged and neutral pions in $p\gamma$ interactions.}
The circles indicate equal energy going into gamma rays and into pairs of muon neutrinos and antineutrinos, which IceCube cannot distinguish between. Because the charged pion energy is shared roughly equally among four particles and the neutral pion energy among two photons, the photons have twice the energy of the neutrinos. Unlike the weakly interacting neutrinos, the gamma rays and electrons may lose energy in the target and will lose additional energy in the background light (EBL) before reaching Earth.}
\label{fig:flow}
\end{figure}

The rationale for searching for cosmic-ray sources by observing neutrinos is straightforward: in relativistic particle flows---for instance, onto black holes---some of the gravitational energy released in the accretion of matter is transformed into the acceleration of protons or heavier nuclei.
These subsequently interact with radiation and/or ambient matter in the vicinity of the black hole to produce pions and other secondary particles that decay into neutrinos.
Both neutrinos and gamma rays are produced with roughly equal rates. While neutral pions decay into two gamma rays, $\pi^0\to\gamma+\gamma$, the charged pions decay into three high-energy neutrinos ($\nu$) and antineutrinos ($\bar\nu$) via the decay chain $\pi^+\to\mu^++\nu_\mu$ followed by $\mu^+\to e^++\bar\nu_\mu +\nu_e$; see~\cref{fig:flow}.
Based on this simplified flow diagram, we expect equal fluxes of gamma rays and muon neutrinos.
The flow diagram of~\cref{fig:flow} implies a multimessenger interface between the pionic gamma-ray flux and the three-flavor neutrino flux:
\begin{equation}\label{eqn:E2dNdE_gamma_nu}
    E_\gamma^2 \frac{dN_{\gamma}}{dE_\gamma} \approx \frac{4}{K_\pi }\, \frac{1}{3} E_\nu^2 \frac{dN}{dE_\nu}  \Big| _{E_\nu = {E_\gamma}/2},
\end{equation}
where $K_\pi$ is the ratio of charged and neutral pions produced, with $K_\pi \approx 2 (1)$ for $pp (p\gamma$) interactions.

\begin{figure*}[t]
\centering
\includegraphics[width=0.7\textwidth]{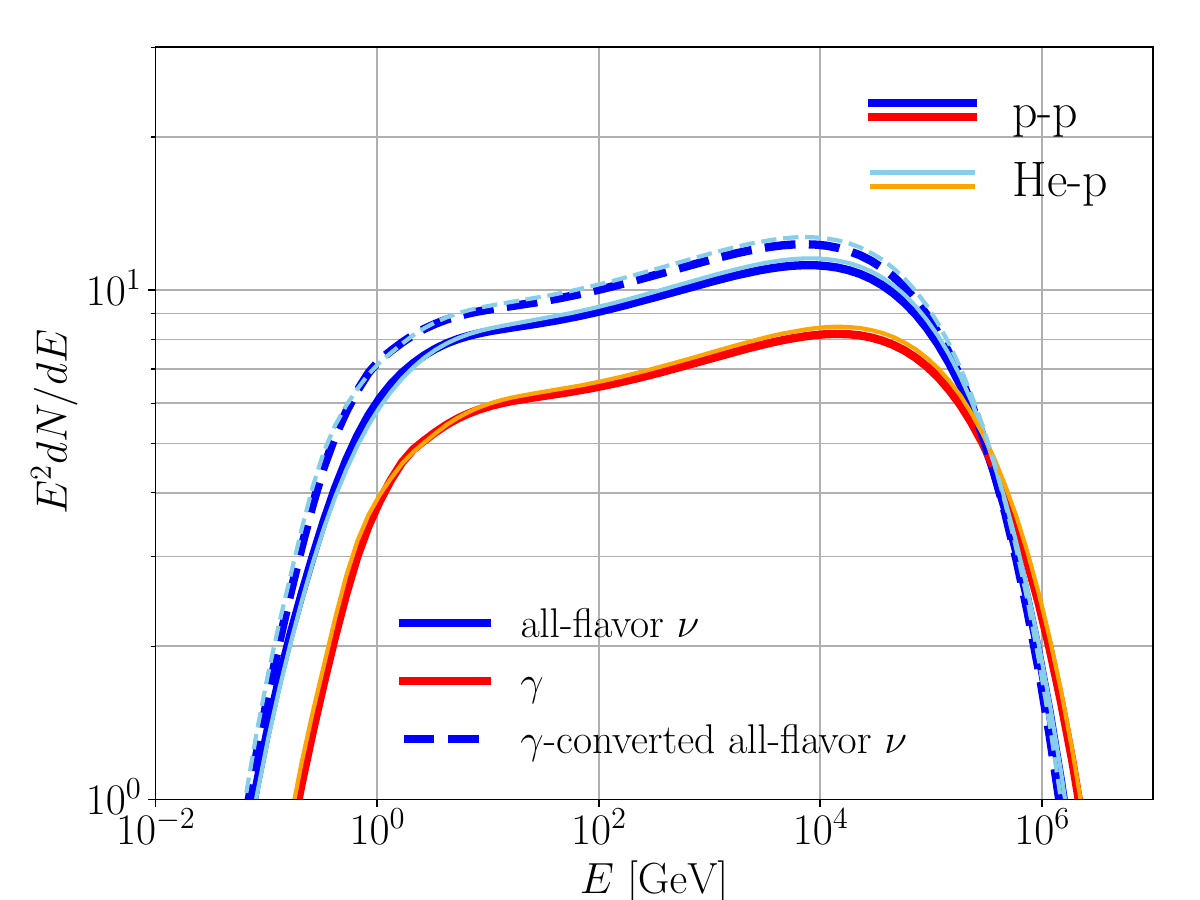}
 \caption{{\bf Energy spectra of secondary gamma rays (red) and neutrinos (blue) from the interaction between proton (dark colors)/helium (light colors) primaries and rest-mass proton targets}. The primary spectrum in use is $dN/dE = E^{-2}\,\exp(-E / 10\,{\rm PeV})$, with $E = E_p = E_{\rm He}/4$ being the energy per nucleon. We have set $E_{\rm min} = 2.5\,\rm GeV$. The secondary spectra are computed using the \texttt{aafragpy} package \cite{Koldobskiy:2021nld} with cross sections from \cite{2006ApJ...647..692K, Kafexhiu:2014cua,Kachelriess:2015wpa}.  For comparison, the dashed curves show the neutrino flux converted from the gamma-ray flux using Eq.~\ref{eqn:E2dNdE_gamma_nu}.} 
\label{fig:ppConversion} 
\end{figure*}

~\Cref{fig:ppConversion} verifies this relation for $pp$ and He-$p$ interactions. We have shown that 1) the conversion between secondary neutrinos and gamma rays expressed in Eq.~\ref{eqn:E2dNdE_gamma_nu} is accurate up to $\sim 10\%$ level, and 2) in a hadronuclear interaction, the chemical composition of the primary cosmic rays barely matters as the $\gamma-\nu$ relation only relies on the production of pions \cite{Fang:2014qva}. 
 
This powerful relation connects neutrinos and pionic gamma rays with no reference to the cosmic-ray beam producing the neutrinos; it simply reflects the fact that a $\pi^0$ decays into two gamma rays for every charged pion producing a $\nu_\mu +\bar\nu_\mu$ pair. 
Also, from the fact that in the photoproduction process 20\% of the initial proton energy is transferred to the pion\footnote{This is referred to as the inelasticity $\kappa_{p\gamma} \simeq 0.2$.}, we anticipate that the gamma ray carries one-tenth of the proton energy and the neutrino approximately half of that. Because cosmic neutrinos are inevitably accompanied by high-energy photons, neutrino astronomy is a multimessenger astronomy.

\begin{figure}[ht]
\centering
\includegraphics[width=0.75\linewidth, trim=0 50 0 80, clip=true]{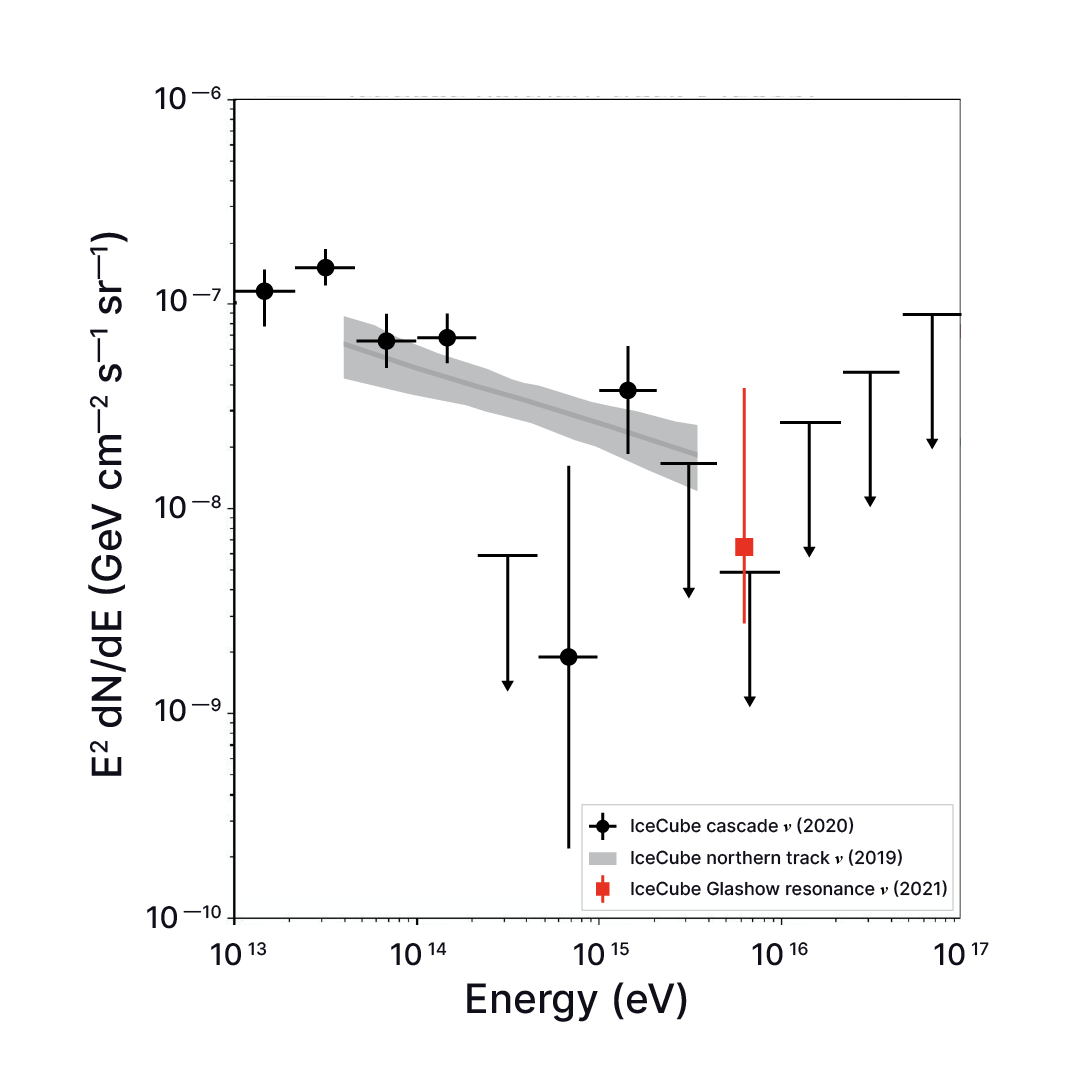}
\caption{\textbf{Current all-sky measurements of the high-energy astrophysical neutrino emission.} 
The flux of cosmic muon neutrinos~\cite{Aartsen:2017mau} inferred from the $9.5$-year upgoing-muon track analysis (solid line) with $1\sigma$ uncertainty range (shaded) is compared with the flux of showers initiated by electron and tau neutrinos~\cite{IceCube:2020acn}, when assuming standard oscillations.
The measurements are consistent with the expectation that each neutrino flavor contributes an identical flux to the diffuse spectrum.
Both are consistent with the flux derived from the observation of a single Glashow-resonance event, shown in red.}
\label{fig:showerstracks-2}
\end{figure}

After accumulating a decade of data, the observed flux of neutrinos reaching us from the cosmos is shown in~\cref{fig:showerstracks-2}.
It has been measured using two different methods to separate the high-energy cosmic neutrinos from the large backgrounds of cosmic-ray muons, 3,000 per second, and neutrinos, one every few minutes, produced by cosmic rays in the atmosphere.
The first method identifies muon neutrinos of cosmic origin from tracks that are produced by upgoing muon neutrinos reaching the South Pole from the Northern Hemisphere.
The Earth is used as a passive shield for the large background of cosmic-ray muons.
By separating the flux of high-energy cosmic neutrinos from the lower energy flux of neutrinos of atmospheric origin, a cosmic high-energy component is isolated with a spectral index $\rm dN/dE \sim E^{-\gamma}$ with $\gamma = 2.37\pm0.09$ above an energy of $\sim 100$\,TeV~\cite{Aartsen:2017mau}; see~\cref{fig:showerstracks-2}.
Also shown are the results of a second search that exclusively identifies showers initiated by electron and tau neutrinos that interact inside the detector and can be isolated from the atmospheric background to energies as low as 10\,TeV~\cite{IceCube:2020fpi}.
Reaching us over long baselines, the cosmic neutrinos have oscillated to a $\nu_e:\nu_\tau:\nu_\mu$ ratio of approximately 1:1:1.

IceCube has independently confirmed the existence of neutrinos of cosmic origin by the observation of high-energy tau neutrinos~\cite{Abbasi:2020zmr} and by the identification of a Glashow-resonance event where a weak intermediate $W^-$ boson is produced in the resonant interaction of an electron antineutrino with an atomic electron: $\bar{\nu}_e + e^- \rightarrow W^- \rightarrow q + \bar{q}$~\cite{IceCube:2021rpz}.
Found in a dedicated search for partially contained showers of very high energy, the reconstructed energy of the Glashow event is 6.3\,PeV, matching the laboratory energy to produce a $W$ boson of mass $80.37$~GeV. Given its high energy, the initial neutrino is deemed cosmic in origin and provides an independent discovery of cosmic neutrinos at the level of $5.2\sigma$.

\begin{figure*}[t]
\centering
\includegraphics[width=0.85\textwidth]{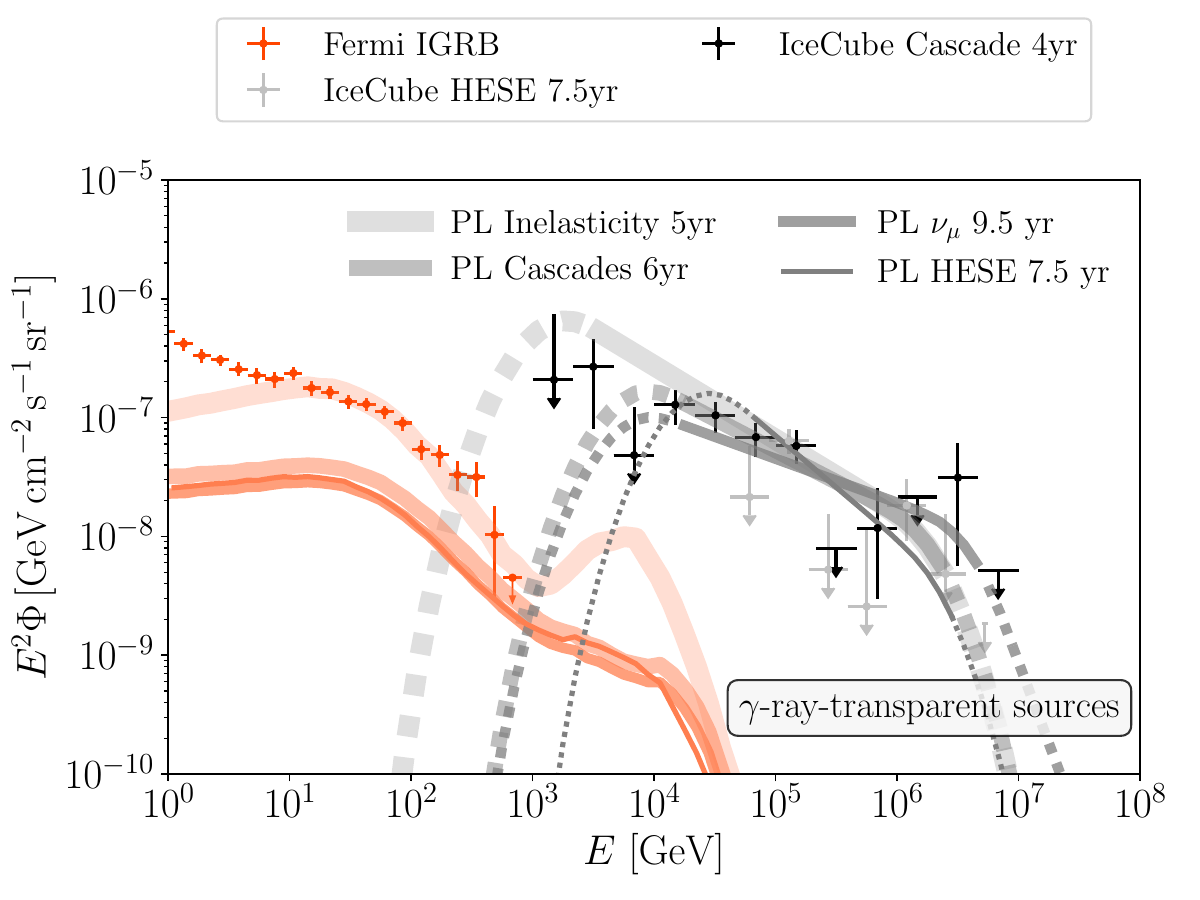}
\caption{\textbf{Gamma-ray-transparent sources?} Cosmic neutrino spectra (gray curves) and gamma-ray flux resulting from the accompanying gamma rays (red curves). The data points are measurements of the diffuse cosmic neutrino flux \citep{2017arXiv171001191I, HESE75yr} and the isotropic gamma-ray background (IGRB)~\citep{FermiIGRB}. Neutrino spectra correspond to the best-fit single power-law models for the different IceCube analyses. Fluxes below and above the sensitivity range for the IceCube analyses are not constrained by the data and are shown as dashed curves.  Pionic gamma rays from hadronic interactions are assumed to leave the sources without attenuation, propagate in the EBL, and cascade down to GeV-TeV energies.} 
\label{fig:seda} 
\end{figure*}

Unlike neutrinos, gamma rays interact with photons associated with the extragalactic background light (EBL) while propagating to Earth.
The gamma-rays lose energy by $e^+e^-$ pair production and the resulting electromagnetic shower subdivides the initial photon energy into multiple photons of reduced energy reaching Earth.
This significantly modifies the implications of the gamma-neutrino interface, which we illustrate~\cite{Fang:2022trf} by first parametrizing the neutrino spectrum as a power law, 
\begin{equation}
  \frac{dN}{dE_\nu}  \propto E_\nu^{-\gamma_{\rm astro}},\, E_{\nu, \rm min} \leq E_\nu \leq E_{\nu, \rm max},\,
\label{eq:powerlaw}
\end{equation}
for four recent measurements taken from Refs.~\cite{HESE75yr, cascade6yr, numu95yr, inelasticity5yr}.
Conservatively, the minimum and maximum energies are limited to the range of neutrino energies that the particular analysis is sensitive to.
For each neutrino flux measurement, we derive the accompanying gamma-ray flux from the fit to the neutrino spectrum using the interface of~\cref{eq:powerlaw}.
This pionic gamma-ray flux is subsequently injected into the EBL assuming that the sources follow the star-formation (SFR) history in a flat $\Lambda$CDM universe \citep{2006ApJ...651..142H}.
\Cref{fig:seda} shows the diffuse neutrino fluxes and those of their accompanying gamma-ray counterparts.
As the values of the neutrino flux spectral indices are comparable for the four measurements, the flux of the cascades is essentially determined by the value of $E_{\nu,\rm min}$ where the energy peaks.
We conclude that when $E_{\nu,\rm min}\lesssim 10$~TeV, the showers initiated by the pionic photons contribute up to $30 \sim 50\%$ of the diffuse extragalactic gamma-ray flux between 30~GeV and 300~GeV and nearly 100\% above 500~GeV.
The cascade flux exceeds the observed gamma-ray flux above $\sim 10$~GeV when $E_{\nu,\rm min} \leq 10$~TeV, assuming that the measured power-law distribution continues.

In the end, the room left to accommodate secondary photons from TeV-PeV gamma rays and electrons is small, and it is likely that the extragalactic gamma-ray flux accompanying cosmic neutrinos exceeds the diffuse gamma-ray flux observed by {\it Fermi}, especially because $\sim 80\%$ of the latter is produced by blazars at the highest gamma-ray energies that are not necessarily sources of neutrinos~\cite{IceCube:2016qvd}. 
There is no contradiction here; we infer from the calculations that the pionic gamma rays already lose energy in the target producing the neutrinos prior to propagating in the EBL.
As a result, pionic gamma rays emerge below the {\it Fermi} threshold, at MeV energies or below.
The observed diffuse neutrino flux originates from gamma-ray-obscured sources. The analysis presented here reinforces the conclusions of previous analyses; see Refs.~\cite{2013PhRvD..88l1301M,2016PhRvL.116g1101M, 2020PhRvD.101j3012C, 2021JCAP...02..037C}.

We should emphasize that the fact that powerful neutrino sources are gamma-ray-obscured should not come as a surprise.
The photon and proton opacities in a neutrino-producing target are related by their cross sections (up to a kinematic factor associated with the different thresholds of the two interactions~\cite{10.1093/mnras/227.2.403}),
\begin{equation}
\tau_{\gamma\gamma} \simeq \frac{\sigma_{\gamma\gamma}}{\kappa_{p\gamma} \sigma_{p\gamma}} \, \tau_{p\gamma} \simeq 10^3\,  \tau_{p\gamma}\,,
\label{opacitygamma}
\end{equation}
where $\kappa_{p\gamma} \sim 0.2$ is the inelasticity in $p\gamma$ interactions.
For instance, we should not expect neutrinos to be significantly produced in blazar jets that are transparent to very high-energy gamma rays.
In contrast, the highly obscured dense cores close to supermassive black holes in active galaxies represent an excellent opportunity to produce neutrinos, besides providing opportunities for accelerating protons.

Recently, IceCube has identified the subdominant flux in the neutrino sky map emitted by the Milky Way. While of importance for resolving the origin of the high-energy radiation from our own galaxy, the discovery also creates new opportunities to search for the still unknown sources of Galactic cosmic rays.

\section{The Discovery of Neutrinos from the Galaxy}

Observing the neutrino emission from our own galaxy has been a challenge because, surprisingly, it is not a prominent feature of the neutrino sky, which is dominated by extragalactic sources. Even after combining IceCube and ANTARES data~\cite{IceCube:2017trr,IceCube:2019lzm,ANTARES:2018nyb,ANTARES:2017nlh}, no statistically significant signal was found, achieving p-values of $\lesssim 0.02$ only. The advent of the Pass-2 data delivering a decade of superior IceCube data, and the rapid development of deep learning techniques in data science produced new opportunities to search for a Galactic component in the flux of cosmic neutrinos~\cite{Abbasi:2021ryj,IceCube:2021umt,mirco_huennefeld_2022_7412035}. Modern neural networks allowed us to identify a larger number of neutrinos in the presence of the dominant atmospheric backgrounds and to reconstruct them with improved angular resolution. A new sample of more than 60,000 shower events achieving lower energies, improved reconstruction by a factor of two, and higher statistics by a factor of twenty made the identification of the Milky Way in the neutrino sky map possible.

Neutrino emission from the Galactic Center and the bulk of the Galactic plane originates in the southern sky, where atmospheric muons represent a challenging background for identifying neutrinos; IceCube records 100 million muons for every astrophysical neutrino observed. IceCube's view of the Southern Hemisphere, therefore, relies on shower events starting inside the detector, which reduces the contamination of atmospheric neutrinos by an order of magnitude at TeV energies and lowers the energy threshold for observing neutrinos from more than 100 TeV to about 1 TeV. In summary, in the southern sky, the lower background, better energy resolution, and lower energy threshold of shower events compensate for their inferior angular resolution compared to secondary muon tracks. This is particularly true for searches of emission from extended objects, such as the Galactic plane.

Here, as with any search for sources in the neutrino sky map, IceCube uses the maximum likelihood method, which yields p-values that include all trials. Using this method~\cite{Braun:2008bg}, an unbinned maximum likelihood is constructed to search for spatial clustering of neutrino events. The significance of an observation of clustering is estimated by confronting the hypothesis of a source with the hypothesis that it is a fluctuation of the atmospheric background. For an event with reconstructed direction $\Vec{x}_i = (\alpha_i, \delta_i)$ representing the declination and right ascension, the probability of originating from the source at $x_s$ is modeled as a circular two-dimensional Gaussian. The signal probability distribution function (PDF)  $\mathcal{S}_i$ incorporates the directional uncertainty $\sigma_i$ and the energy $E_i$ for each individual event and the angular difference between the reconstructed direction of the event and the source:
\begin{equation}\label{SPDF}
\mathcal{S}_i ={S}_i(|\Vec{x}_i - \Vec{x}_s|, \sigma_i)\, \mathcal{E}_i(\delta_i, E_i, \gamma),
\end{equation}
where the spatial distribution is modeled as a two-dimensional Gaussian 
\begin{equation}
S_i(|\Vec{x}_i - \Vec{x}_s|, \sigma_i) = \frac{1}{2\pi \sigma_i^2} \exp \Big( {-\frac{|\Vec{x}_i - \Vec{x}_s|^2}{2\sigma_i^2}}\Big)
\end{equation}
and $\gamma$ is the spectral index. The energy distribution $\mathcal{E}_i(\delta_i, E_i, \gamma)$ represents the probability of observing the reconstructed muon energy $E_i$ given the source spectral index $\gamma$ assuming a power-law energy spectrum. A harder spectrum relative to the background can also indicate a signal.

The background PDF, $\mathcal{B}_i$, is assumed to be atmospheric in origin and can be calculated. Here we assume that the data in each declination/zenith range are background-dominated and the total event rate is a good approximation to the background.

The likelihood function for a point source is defined as the sum of events $N$ originating from signal and background: 
\begin{equation}
\mathcal{L}(n_s,x_s, \gamma) = \prod_i^{events}\bigg( \frac{n_s}{N} \mathcal{S}_i(|x_i - x_s|,\sigma_i, E_i, \gamma) + \frac{N - n_s}{N} \mathcal{B}_i(\delta_i, E_i) \bigg).
\end{equation}

After determining the best fit for the number of signal events $\hat n_s$ and their spectral index $\hat \gamma$, the TS is calculated as
\begin{equation}
{\rm TS} = 2 \log \big[ \frac{\mathcal{L}(\hat n_s,\hat \gamma)}{\mathcal{L}(n_s=0)} \big].
\end{equation}
The significance of an observation is determined by comparing the maximized TS to the TS distribution of the background, usually obtained from data randomized in right ascension. For large sample sizes, this distribution approximately follows a chi-squared distribution, where the number of degrees of freedom corresponds to the difference in the number of free parameters between the null hypothesis and the alternate hypothesis.

The search can be modified to search for extended sources of neutrinos. For this purpose, the angular uncertainty in the spatial PDF is modified to include the extension of the source:
\begin{equation}
\sigma_{\rm eff} =  \sqrt{\sigma_i^2+\sigma_{\rm ext}^2}.
\end{equation}
The technique described above can be extended to any hypothesis that may represent multiple sources or any extended feature in the sky, such as the Galactic plane.

IceCube constructed a full-sky template of the Galactic plane using the GeV-energy observations by {\it Fermi}. Assuming that the observed gamma-ray flux consists of photons that are the decay products of pions produced by Galactic cosmic rays colliding with the interstellar medium, it can be transformed into a neutrino template for the Galactic plane using the multimessenger interface described above. This template is subsequently binned in equal-solid-angle bins, convolved with the detector acceptance, and smeared with a Gaussian with a width equal to each event’s angular uncertainty. Each event, therefore, has a particular spatial weight based on its position, energy, direction, estimated angular uncertainty, and the signal hypothesis under investigation. These combined spatial and energy weights are multiplied and the likelihood maximized for the analysis parameters described above.

\begin{figure*}[ht!]
\centering
\includegraphics[width=0.85\textwidth]{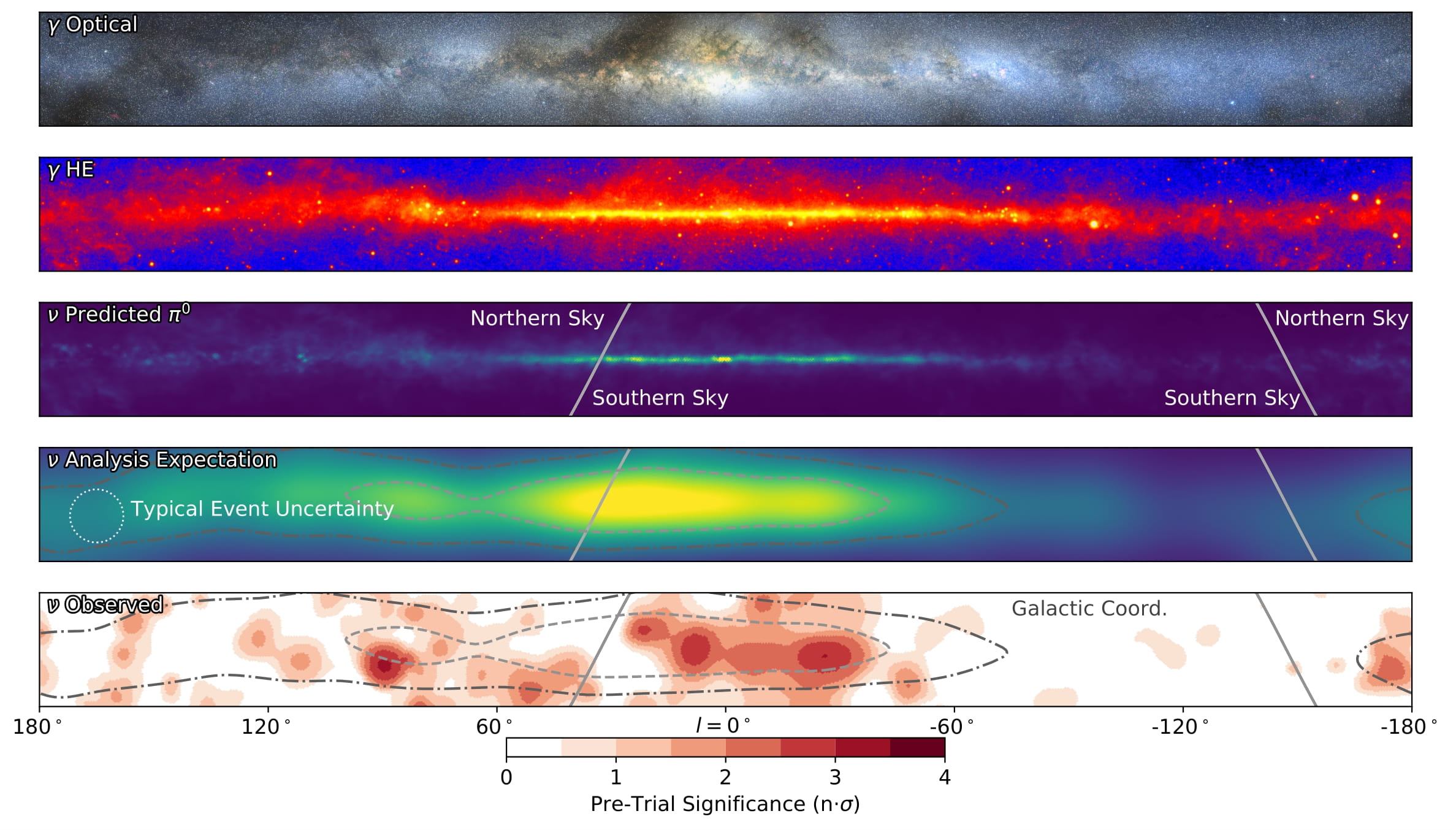}
\caption{{\bf The plane of the Milky Way galaxy in photons and neutrinos.} Each panel from top to bottom shows the flux in latitude and longitude: (A) Optical flux. (B) The integrated flux in gamma rays from the Fermi-LAT 12-year survey at energies greater than 1 GeV. (C) The neutrino emission template derived from the $\pi^0$ template that matches the {\it Fermi}-LAT observations of the diffuse gamma-ray emission. (D) The emission template from panel (C) including the IceCube detector sensitivity to cascade-like neutrino events and the angular uncertainty of a typical signal event (7 degrees, indicated by the dotted white circle). Contours indicate the central regions that contain 20\% and 50\% of the predicted diffuse neutrino emission signal. (E) The pre-trial significance of the IceCube neutrino observations, calculated from the all-sky scan for point-like sources using the cascade neutrino event sample. Contours are the same as in panel (D). For more details, see Ref.~\cite{IceCubeGP}.} 
\label{fig:MultimessengerGalacticPlane} 
\end{figure*}

The construction of the template requires modeling the {\it Fermi} data. Three such sky templates were used in the search using cascades. The first is a GALPROP model \cite{Strong:1998pw} calibrated with the {\it Fermi} observation, referred to as the $\pi^0$ template and illustrated in~\cref{fig:MultimessengerGalacticPlane}. The other two are KRA models that include radially dependent cosmic-ray transport properties. In particular, IceCube used the KRA models with a cosmic-ray cut-off at 5 and 50~PeV \cite{Gaggero:2015xza}, denoted as the KRA$_\gamma^{5}$ and KRA$_\gamma^{50}$ models, respectively.  In addition to these three templates, the IceCube search using tracks \cite{IceCube:2023hou} used the CRINGE template, based on a global fit to cosmic-ray observations \cite{Schwefer:2022zly} and analytical models from \cite{Fang:2021ylv}. 

\begin{figure*}[t]
\centering
\includegraphics[width=0.65\textwidth]{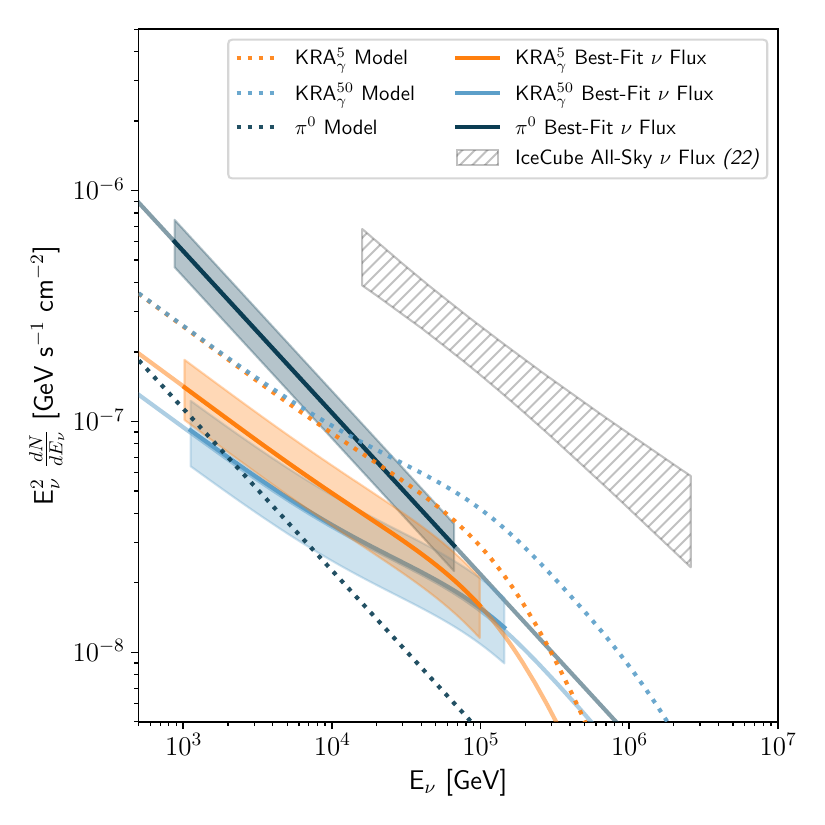}
\caption{{\bf The energy spectra of neutrinos from the Galactic plane measured using three templates of the diffuse neutrino emission: the $\pi^0$, KRA$_\gamma^{5}$ and KRA$_\gamma^{50}$ models} (from \cite{IceCube:2023ame}). Dotted curves indicate the model flux while solid curves and shaded regions indicate the measured flux and  $1\sigma$ uncertainties, respectively. The hatched region shows the IceCube all-sky diffuse flux \cite{Aartsen:2020aqd}.  } 
\label{fig:diffuseNu} 
\end{figure*}

The Galactic template search was performed with the same methods as previous Galactic diffuse emission searches~\cite{IceCube:2017trr,IceCube:2019lzm}. The analysis has only one unconstrained parameter, the number of signal events $n_s$ in the entire sky, providing the flux normalization while keeping the spectrum fixed to the template. The search with cascades rejects the background-only hypothesis with a significance of $4.71\sigma$, which is reduced to $4.5\sigma$ because of trials related to other searches performed; for details on this and the maximum likelihood method, we refer to the supplementary material in Ref.~\cite{IceCube:2023ame}. The result is shown in ~\cref{fig:diffuseNu}. The search with tracks identified a Galactic contribution in the northern sky at the $2.7\sigma$ level with a consistent flux.

The ANTARES collaboration searched for neutrino emission from the Galactic ridge region, defined as $|l|<30^\circ$ and $|b|<2^\circ$ \cite{ANTARES:2022izu}. By comparing the flux of neutrino events in the ``on" region, i.e., the Galactic ridge, to that in a off-zone region and describing the neutrino spectrum as a power law, Ref.~\cite{ANTARES:2022izu} identified a mild excess of neutrino emission from the Galactic ridge region at the level of $2.2\,\sigma$. The measured neutrino flux is consistent with the model of diffuse gamma rays from the $\pi^0$ decays from the interaction of protons with the ISM in the Galactic ridge. The best-fit spectral index of neutrinos is $dN/dE_\nu\propto E_\nu^{-2.45}$, suggesting a harder cosmic-ray spectrum in the inner Galaxy. 

%\section{Modeling the Neutrino Flux from the Milky Way of Cosmic Ray Origin}

\section{Modeling the Diffuse Neutrino Flux from the Galactic Plane}

%The consistency suggests that the diffuse $\gamma$-ray emission from our Galaxy above $\sim$1~TeV is dominated by the production and decay of neutral and charged pions. The template used to extract the Galactic component from the neutrino observations was indeed constructed on the assumption that below $\sim 10$~TeV the flux measured by the {\it Fermi} Large Area Telescope (LAT) \cite{Fermi-LAT:2012edv} is of $\pi^0$ origin as discussed above.

The Galactic diffuse emission (GDE) in gamma rays (GDE-$\gamma$) has been measured by {\it Fermi}-LAT from 100~MeV to 1~TeV~\cite{Fermi-LAT:2012edv}. It has been observed by ground-based gamma-ray experiments above $\sim$1~TeV~\cite{HESS:2014ree, HAWC:2021bvb}, from the parts of the Galactic plane that are accessible to the detectors, and above 100~TeV by the Tibet AS$\gamma$~\cite{Tibet21}, HAWC~\cite{HAWC:2021bvb}, and LHAASO-KM2A~\cite{lhaasoGDE} observatories. Most of the photons observed by AS$\gamma$ with energy above 398~TeV do not point to known gamma-ray sources, indicating that the emission could be diffuse in nature \citep{Kato:2024kxn}. On the other hand, LHAASO-KM2A finds a lower GDE intensity for a similar sky region when masking known sources detected by LHAASO \cite{lhaasoGDE}. The comparison suggests that a significant fraction of the AS$\gamma$ flux could be contributed by unresolved sources \cite{2022ApJ...928...19V}. 

\begin{figure*}[t]
\centering
%\includegraphics[width=0.79\textwidth]{figures/galactic_diffuse.png}
%trim={<left> <lower> <right> <upper>}
\includegraphics[width=0.79\textwidth, trim={{.02\textwidth} {.045\textwidth} 0 {.05\textwidth}},clip]{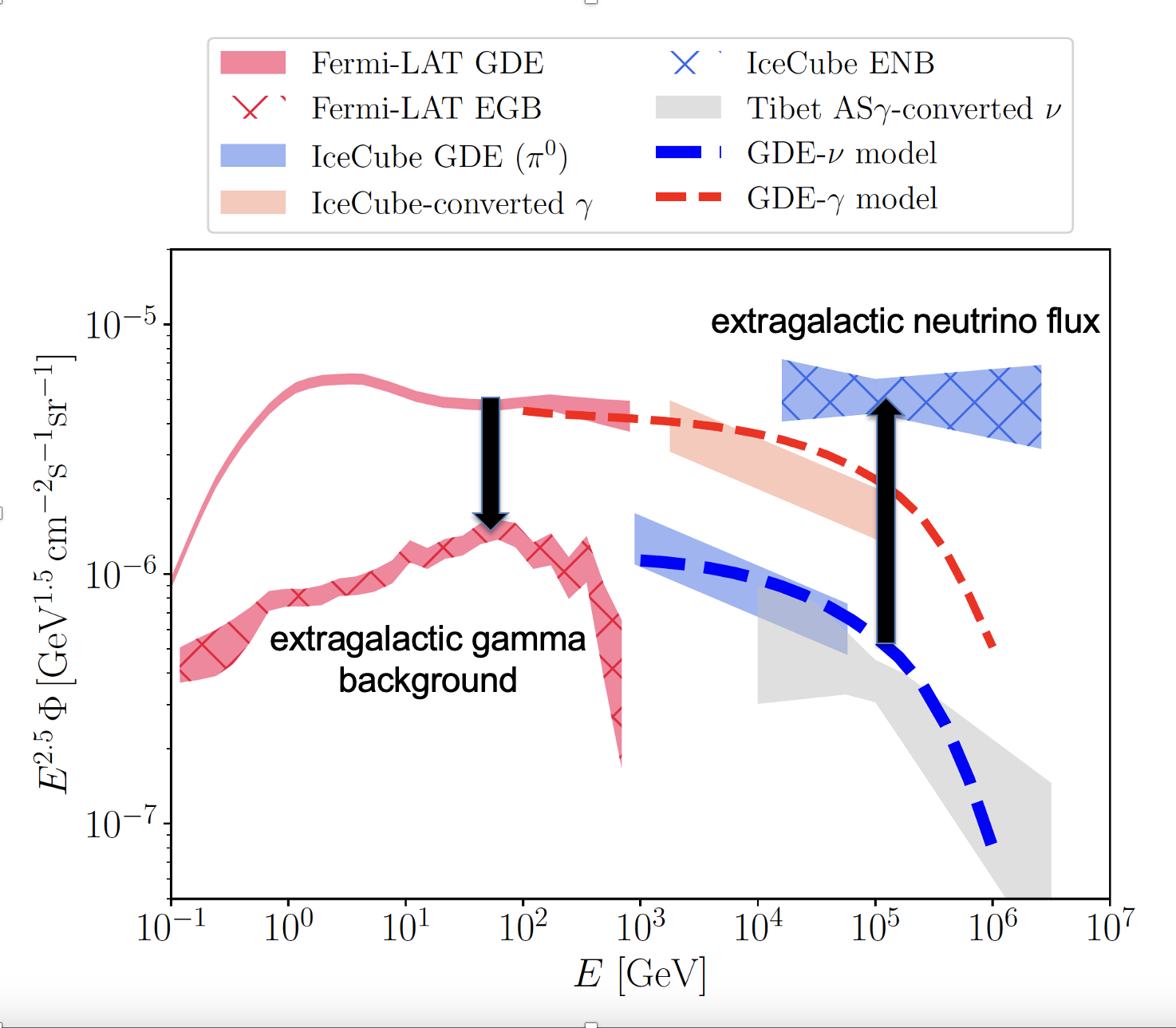}
\caption{{\bf The neutrino flux of the Galactic plane.} The neutrino flux is shown in blue with a spread that includes statistical and systematic errors. The accompanying gamma-ray flux shown in red matches the lower energy Fermi data. Both are consistent with a model (dashed lines) determining the number of parent pions produced in interactions of Galactic cosmic rays interacting with the interstellar medium. The neutrino flux observed by IceCube is consistent with the one observed by the AS$\gamma$ array at the higher energies after its conversion to neutrinos~\cite{Tibet21}. The diffuse extragalactic gamma-ray (red hatched) and neutrino flux (blue hatched) are shown for comparison.} 
\label{fig:diffuse} 
\end{figure*}

The flux of pionic gamma rays that accompanies the flux of high-energy neutrinos from the Galactic plane is consistent with the Galactic diffuse gamma-ray emission observed by the {\it Fermi}-LAT and AS$\gamma$ experiments around 1~TeV and 0.5~PeV, respectively; see~\cref{fig:diffuse}. This consistency suggests that the GDE-$\gamma$ could be dominantly produced by hadronic interaction above $\sim$1~TeV. In addition, the IceCube flux is comparable to the sum of the flux of individual sources not associated with pulsars and the LHAASO diffuse emission above $\sim$30 TeV (as shown in Figure~4 of \cite{Fang:2023ffx}). This implies that the LHAASO GDE and a significant portion of the resolved and unresolved sources may dominantly originate from hadronic interactions. 

A priori, the IceCube Galactic plane flux may be explained as:

\begin{enumerate}
  \item Unresolved emission by individual neutrino sources. 
  \item The truly diffuse emission produced when cosmic rays confined by the Galactic magnetic field, the ``cosmic-ray sea,'' interact with the interstellar medium. 
  \item A combination of both components. 
\end{enumerate}
 
In Scenario 1, comparing the neutrino flux converted from the gamma-ray flux of Galactic sources observed by various imaging atmospheric Cherenkov telescopes (IACTs) and air shower gamma-ray detectors with the IceCube results found that the gamma-ray sources are insufficient to accommodate the IceCube Galactic plane flux, especially considering that gamma rays may also be produced by electrons \cite{Fang:2023ffx, 2024arXiv240305288G, 2024NatAs.tmp...54Y}. Therefore, sources that are currently unresolved by gamma-ray facilities need to be present to explain the IceCube observation. Particularly interesting candidate sources include young massive star clusters \cite{2024PhRvD.109d3007A}, hypernova remnants \cite{Ahlers:2013xia, Fang:2021ylv}, and giant molecular clouds \cite{2024arXiv240105863R}. 

Scenario 2 was long predicted by Refs.~\cite{Gaggero:2015xza, Lipari:2018gzn, Fang:2021ylv, 2022A&A...661A..72B, 2023A&A...672A..58D}, for example, and more recently employed by Refs.~\cite{Fang:2023azx,2023JCAP...09..027V, 2023ApJ...956L..44V} to explain the IceCube observation. Below we present a model illustrating this idea, realizing at the same time that given the uncertainty of the GDE-$\nu$ flux associated with the analysis templates and the potential contribution from unresolved neutrino sources, leptonic processes may still play a role, in particular between 0.1 and a few TeV.

We first model the cosmic-ray density in the Galaxy and subsequently calculate the production of photons and neutrinos in the interactions of cosmic rays with the interstellar gas.

% cosmic-ray knee and PeVatrons; GDE 
%
\begin{figure*}[t]
\centering
\includegraphics[width=0.79\textwidth]{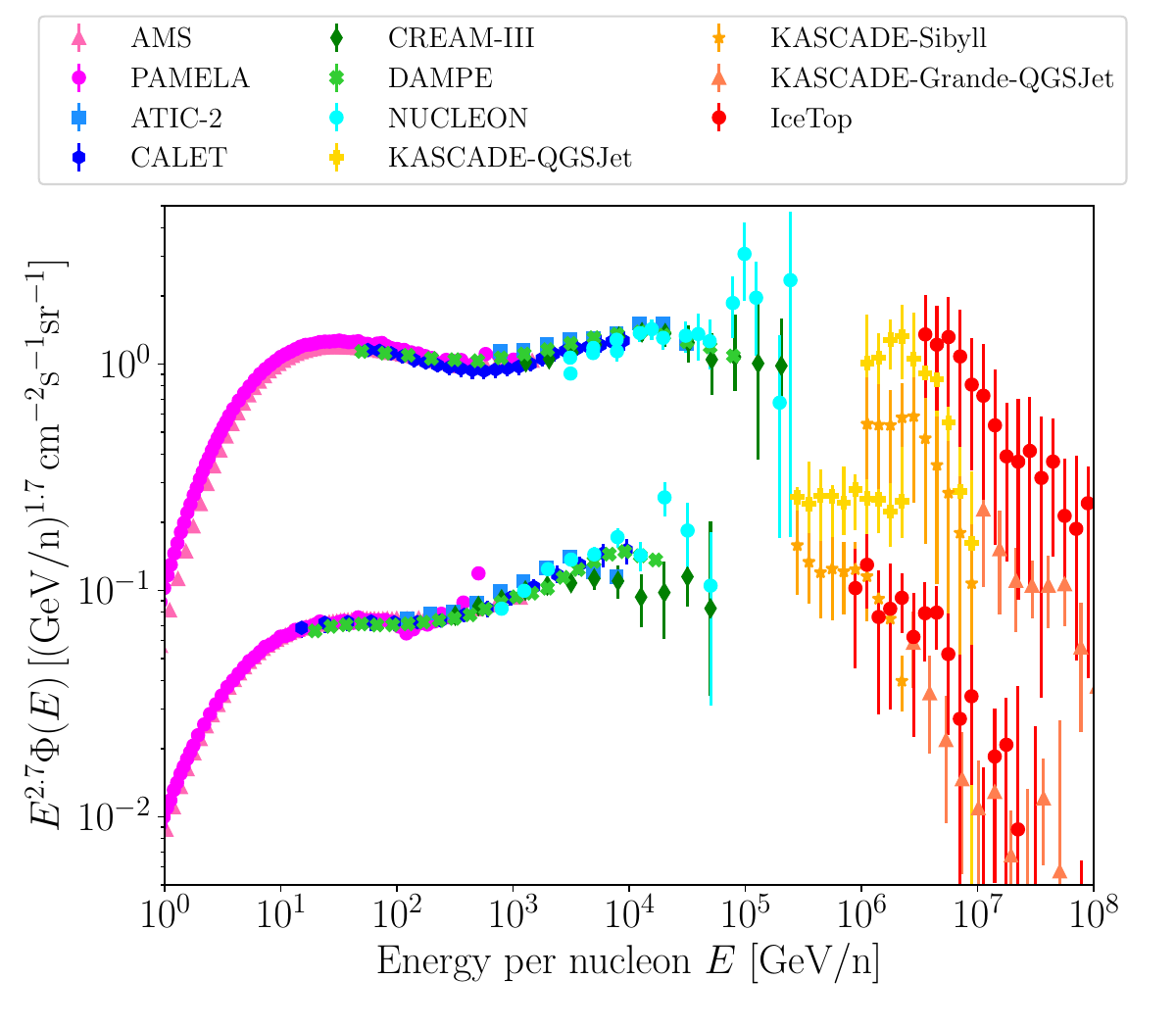}
 \caption{{\bf Cosmic-ray proton (top curve) and helium (bottom curve) fluxes at the Earth between 1~GeV and $10^8$~GeV scaled by $E^{2.7}$} (adapted from Ref. \cite{Fang:2023azx}). The data points are from measurements by AMS-02 \citep{PhysRevLett.114.171103, PhysRevLett.115.211101}, PAMELA \citep{2011Sci...332...69A}, ATIC \citep{2009BRASP..73..564P}, CALET \citep{2019PhRvL.122r1102A, Brogi:2021iB}, CREAM-III \citep{2017ApJ...839....5Y}, DAMPE \citep{2019SciA....5.3793A}, NUCLEON \citep{Atkin:2017vhi}, KASCADE \citep{2013APh....47...54A}, and IceTop \citep{IceCube:2019hmk} experiments. The errors indicate the quadratic sums of statistical and systematic uncertainties of the measurements.  } 
\label{fig:crSED} 
\end{figure*}

In the energy range of interest, Galactic cosmic rays are dominated by proton and helium nulcei. Their spectral shape can be described by a modified power law~\citep{PhysRevLett.114.171103} that cuts off at low and high energies and describes the break near a rigidity of 300~GV (see~\cref{fig:crSED}). The cosmic-ray density in the Galaxy is computed using the public simulation framework CRPropa~3.1~\citep{2017JCAP...06..046M}. Stochastic differential equations are solved to describe particle propagation in the Galactic magnetic field described by the JF12 model, including both the regular and turbulent components \citep{Jansson_2012}. The sources are assumed to follow the distribution of isolated radio pulsars~\citep{2006ApJ...643..332F, 2012JCAP...01..010B}, which trace the spiral structure of the Galaxy. Modeling the spiral-arm source density produces a higher cosmic-ray energy density than a smooth disk when the propagated intensities for each are similarly normalized to the cosmic-ray data. %A face-on view of the cosmic-ray density on the plane ($b=0^\circ$) is shown in Fig.~\ref{fig:crDensity}.

%\begin{figure} 
%    \centering
%   \includegraphics[width=0.75\textwidth]{figures/test_pulsar.pdf}
%    \caption{\label{fig:crDensity} {\bf Face-on view of the Galactic cosmic-ray density at 100~TeV at $b=0^\circ$.} The color scale indicates the ratio of the number density of cosmic rays at a given position $\vec{x}$ to that of the solar neighborhood $\vec{x}_\odot$. The red star marks the location of the sun.  
%    } 
%\end{figure}
 
The model assumes that the spatial and spectral components of the nucleon flux are independent. This assumption could be oversimplified and does not describe the {\it Fermi}-LAT observations~\citep{Acero:2016qlg}. Variations of the cosmic-ray spectra across the Galaxy also affect the more detailed modeling of the GDE-$\nu$ flux~\citep{2015ApJ...815L..25G}. 
%We included the contribution of leptons to the GDE-$\gamma$ flux, but because it is sub-dominant above $\sim 1$~GeV the details of the modeling of leptons barely impact the results. 
For a detailed description of the model, we refer the reader to the supplementary material of Ref.~\cite{Fang:2023azx}.

\begin{figure*} 
    \centering
    \includegraphics[width=0.49\textwidth]{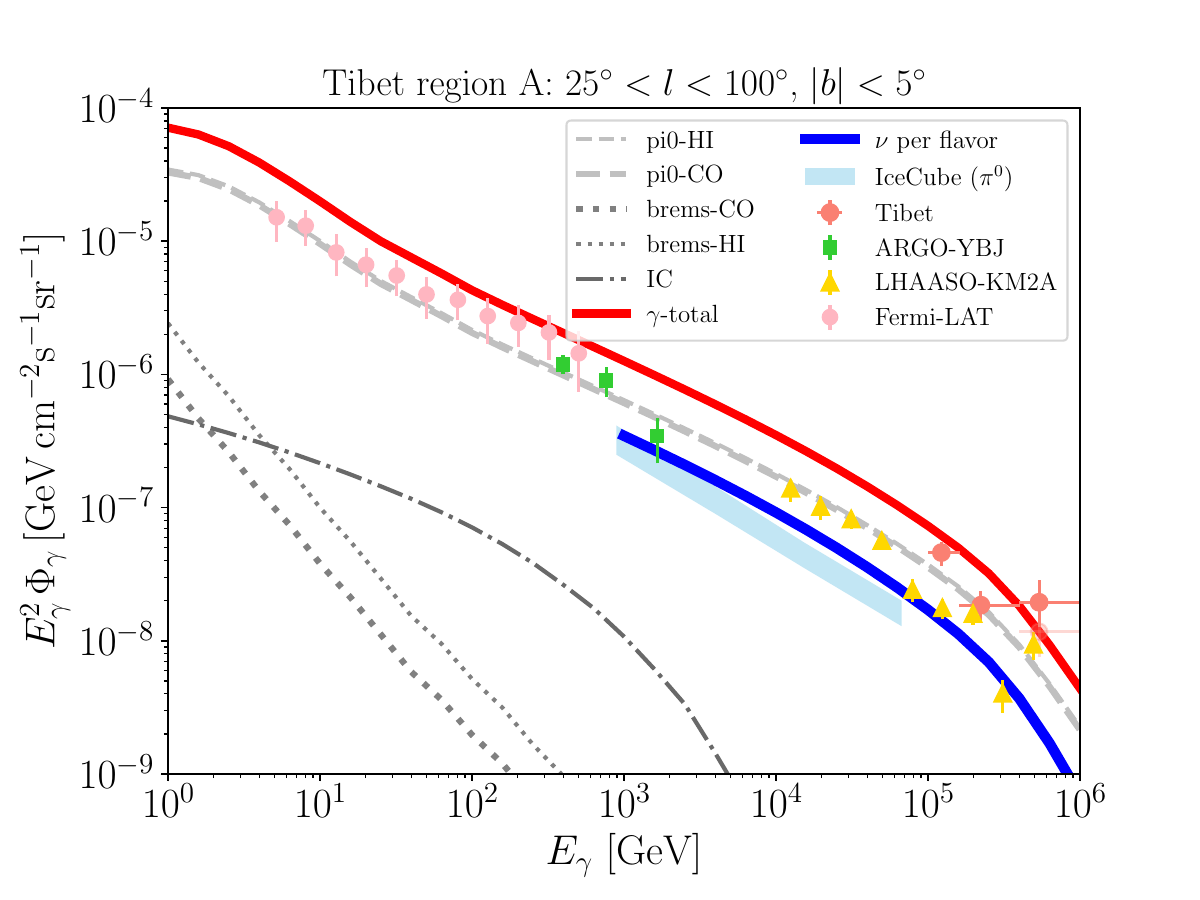}
    \includegraphics[width=0.49\textwidth]{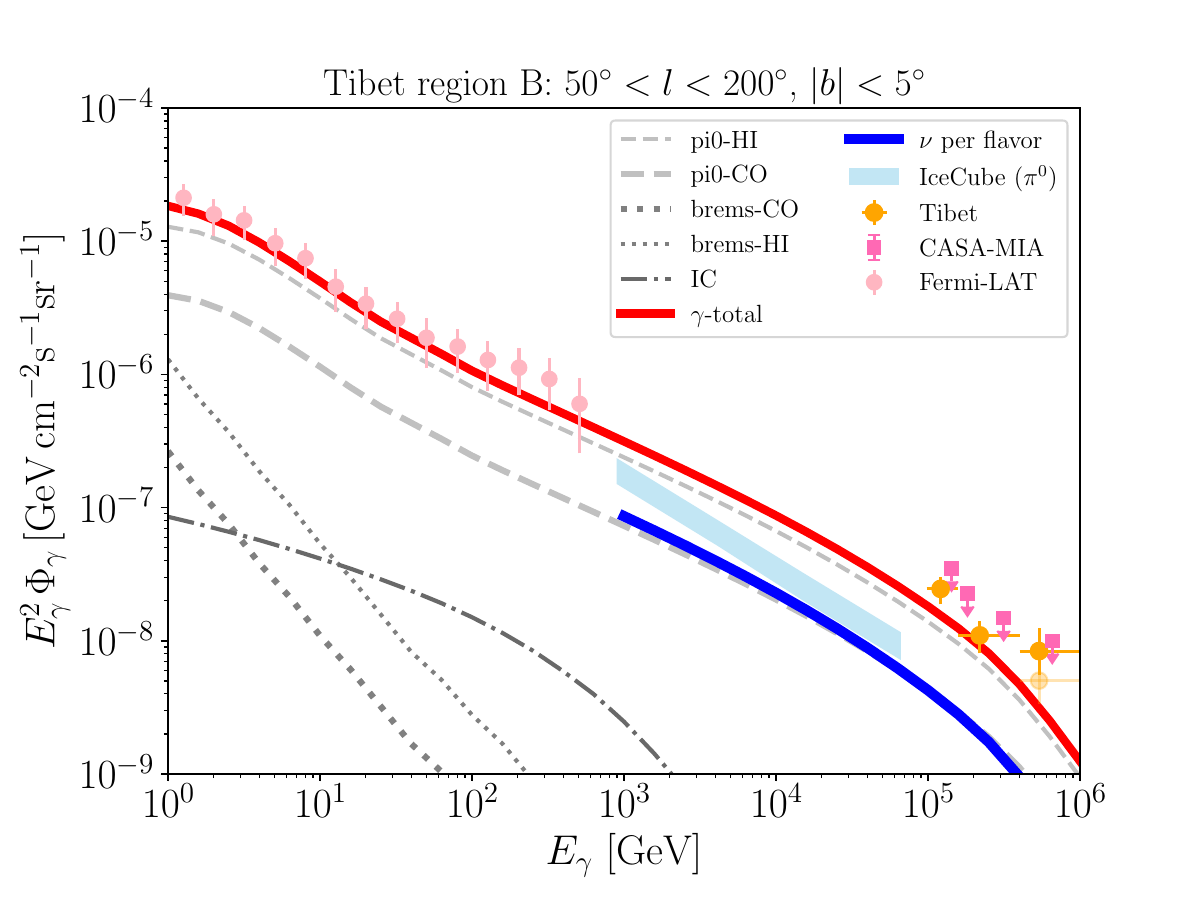}
    \caption{\label{fig:sed} {\bf Intensities of Galactic diffuse emission in gamma rays and neutrinos from two sky regions, {\it region A} (left): $25^\circ < l < 100^\circ$, $|b|<5^\circ$, and {\it region B} (right): $50^\circ < l < 200^\circ$, $|b|<5^\circ$} (from \cite{Fang:2023azx}). The data points refer to  observations of various gamma-ray experiments, including AS$\gamma$ \citep{Tibet21} (orange data points; the fainter data points at 500~TeV indicate the residual intensity after removing events near the Cygnus Cocoon), LHAASO-KM2A \citep{Zhao:2021GJ} (yellow triangle markers), {\it Fermi}-LAT \citep{2022arXiv220315759D} (pink circular markers indicating the data of the corresponding region), ARGO-YBJ \citep{ARGO-YBJ:2015cpa} (green circular markers in {\it region A}), and CASA-MIA \citep{Borione_1998} (magenta square markers in {\it region B}). The error bars show $1\,\sigma$ statistical errors. The gamma-ray contribution (red) is decomposed into $\pi^0$ decay (grey dashed), inverse Compton (grey dash-dotted) and bremsstrahlung emission (grey dotted), with thick and thin curves indicating products from interaction with molecular (H$_2$) and neutral (HI) gas, respectively. The blue curves indicate the corresponding per-flavor neutrino intensities. The gamma-ray and neutrino spectra presented in this figure are from a simple model where the spatial and spectral components of the nucleon flux are assumed to be independent.  
    } 
\end{figure*}

The emissivity and intensity of secondary production is subsequently computed using the \texttt{HERMES} code~\citep{HERMES}, with hadronic interaction cross section from Ref.~\citealp{2006PhRvD..74c4018K} and the local gas density profiles from Refs.~\cite{1998ApJ...497..759F, 2003PASJ...55..191N, 2006PASJ...58..847N}. The calculation accounts for the cosmic-ray interactions with the neutral and molecular hydrogen assumed to have a uniform abundance ratio $n_{\rm He} / n_{\rm H} = 0.1$ traced by the CO emission.  

Our model, shown as the blue dashed curve in~\cref{fig:diffuse}, agrees with the measurements using the $\pi^0$ and CRINGE models~\citep{2023ApJ...949...16S, IceCube:2023hou} but exceed the fluxes predicted by the KRA models~\citep{IceCubeGP}. We note that the model presented in this work is designed to explain the IceCube $\pi^0$ flux as diffuse emission. When changing our source density distribution from a spiral arm model to a smooth disk model, the prediction of the secondary flux would be lowered by a factor of $\sim 2$ and comparable to the $\pi^0$ and KRA models at $\sim 100$~GeV.  

~\Cref{fig:sed} further shows the pionic gamma-ray emission and the inverse Compton and synchrotron radiation of the subdominant electrons in two sky regions well observed by gamma-ray facilities. In {\it region A}, the model accommodates the IceCube and the AS$\gamma$ GDE observations but overproduces the LHAASO-KM2A measurement, which is a factor $\sim$2-3 times lower than the AS$\gamma$ measurement. The difference in multimessenger observations could be attributed to a contribution of hadronic sources that are not resolved by the AS$\gamma$ and IceCube observations. The source contribution, on the other hand, should not alter the conclusion that the hadronic emission dominates the Galactic plane emission above $\sim 30$~TeV, given the agreement between the IceCube and AS$\gamma$ fluxes.  

The longitudinal and latitudinal distributions of our GDE-$\gamma$ model are roughly consistent with {\it Fermi}-LAT observations above 100~GeV. Nevertheless, this simplified model is not tailored to fit their data, known to require multiple emission components~\citep{Acero:2016qlg}. The light blue bands in~\cref{fig:sed} show the average intensity in the two AS$\gamma$ regions of the $\pi^0$ template scaled by the full-sky normalization observed by IceCube. The model produces higher flux in {\it region A} and lower in {\it region B}, suggesting that its spatial distribution differs from the $\pi^0$ template.

The modeling of diffuse neutrino and gamma-ray emission is affected by several poorly known factors, including 1) 
cosmic-ray spectra above the rigidity $\sim$10 TV observed at the Earth, 2) the spatial dependence of the cosmic-ray density, which is determined by the source distribution in the Galaxy, timescales of the sources, and the particle diffusion in the Galactic magnetic field, and 3) the ISM gas density profile. The effects of these factors could be coupled and are not constrained by current observations \cite{2021arXiv211212745P,2017JCAP...02..015E, 2022arXiv220315759D, 2023ApJ...949...16S, Fang:2023azx, 2023ApJ...949...16S}.

%\begin{figure} 
%    \centering
%  \includegraphics[width=0.75\textwidth]{figures/GalSpectraKRA.pdf}
%    \caption{\label{fig:GalSpectraKRA} {\bf Comparison of our Galactic neutrino diffuse emission model with the observed and derived flux of the Galactic Plane.} The blue dashed curve indicates the model. The IceCube observations include measurements using cascade events and the {\it Fermi} $\pi^0$ model (dark blue band), KRA models (green and purple bands) \citep{IceCubeGP} and using track events and the CRINGE model (light blue band) \citep{2023ApJ...949...16S, IceCube:2023hou}. Also shown are the neutrino flux derived from the GDE-$\gamma$ measured by Tibet AS$\gamma$ \citep{Tibet21, FangMurase21} (silver band) and the Galactic interstellar emission model of {\it Fermi}-LAT \citep{4FGL} (turquoise band). All curves except the CRINGE band are $4\pi$-averaged. The IceCube CRINGE flux is based on an analysis with track data from the northern hemisphere and is averaged over the corresponding sensitive sky region.    } 
%\end{figure}

%Finally, Fig.~\ref{fig:GalSpectraKRA} compares the model with various observations of the Galactic plane. Many models of the GDE emission have been developed over the years. 

\section{The Galaxy is a Neutrino Desert}

In \cref{fig:diffuse}, we have also shown the measurements of the extragalactic diffuse flux of photons and neutrinos measured by {\it Fermi} and IceCube, respectively. It is striking how the neutrino flux from the Milky Way is subdominant to the aggregate flux of extragalactic sources. While the Galactic component of the neutrino flux is at the level of 10\% of the extragalactic flux at $30\,\rm TeV$, it is the dominant feature in the gamma-ray flux. This contrast is expressed explicitly by the following relation:
\begin{equation} \label{comparison}
\rm \frac{L_\nu^{EG}}{L_\nu^{MW}} \sim 120\, \left(\frac{\Phi_\nu^{EG}/\Phi_\nu^{MW}}{5}\right)\left(\frac{n_0}{0.01\,Mpc^{-3}}\right)^{-1}\left(\frac{\xi}{3}\right)^{-1}\left(\frac{F_\epsilon}{1}\right)\,,
\end{equation}
which represents the ratio of the neutrino flux from extragalactic sources with a density $n_0$ and luminosity $\rm L^{EG}_\nu$,
\begin{equation} \label{EG}
\Phi^{\rm EG}_\nu = \rm n_0ct_H \,L^{EG}_{\nu}\,\,\,\xi\,,
\end{equation}
and the Galactic neutrino flux with luminosity $\rm L^{MW}_\nu$,
\begin{equation} \label{MW}
\Phi^{\rm MW}_\nu = \rm  \frac{3}{4\pi r_0^2}\,\,\,L_\nu^{MW}\,\,\,F_\epsilon\,.
\end{equation}
In the above equations, $\rm t_H$ is the Hubble time and $r_0$ the distance to the center of the Galaxy; $\xi$ and $\rm F_\epsilon$ are geometric factors of order unity related to the cosmic evolution of the sources and the detailed geometry of our Galaxy, respectively. Eq.\ref{MW} relates the neutrino flux of the Milky Way to the locally observed luminosity.

In Eq.\ref{comparison}, we have assumed the source density for galaxies with a stellar mass similar to that of the Milky Way. It is clear that the extragalactic neutrino flux dominates by two orders of magnitude, even for the conservative assumption of an observed ratio of $\rm \Phi_\nu^{EG}/\Phi_\nu^{MW} \simeq 5$. We conclude that neutrino sources that do not exist in our own galaxy do exist in other galaxies. Given that the initial neutrino sources detected by IceCube are active galaxies, one may speculate that the absence of any major activity of the black hole in our own galaxy, at least for the last few million years, is responsible for the missing neutrinos.

%\section{Multiwavelength Search for Galactic Neutrino Sources}
\section{Multiwavelength Search for Galactic PeVatrons}

The softening of the cosmic-ray spectrum from $dN/dE\propto E^{-2.7}$  to $E^{-3.1}$ at several PeV, shown in~\cref{fig:crSED}, is known as the cosmic-ray ``knee" \cite{2009PrPNP..63..293B}. The presence of the knee suggests that the Galactic cosmic-ray sources accelerate protons up to energies of at least a few PeV, a.k.a. ``PeVatrons." These cosmic rays produce neutrinos of 50-TeV energy and higher when interacting with the ambient matter and photons in the vicinity of their sources, forming individual, and possibly extended, Galactic neutrino sources. Confined by the Galactic magnetic field, these cosmic rays will interact with interstellar gas, forming the Galactic diffuse neutrino emission.  

% signatures of pionic gamma-rays: pion0 decay bump (W44, IC443) and multi-wavelength fits (HESS GC, Cygnus Cocoon, SNR G106) 
Identifying the Galactic sources that accelerate cosmic rays has been one of the major tasks of gamma-ray astronomy \cite{2009ARA&A..47..523H,2012APh....39...61H}. Although the spectral break in the gamma-ray spectrum at sub-GeV energies may present evidence for the acceleration of cosmic rays, it may also be caused by re-acceleration of diffuse cosmic rays for a limited time period. Thus, the signature does not necessarily indicate the acceleration of freshly injected protons \cite{2010ApJ...723L.122U,2016A&A...595A..58C}.

Finding unequivocal evidence for the acceleration of protons through gamma-ray observation has been challenging. Two approaches have been pursued to identify hadronic, also referred to as pionic, gamma-ray emission. 

The first approach is to directly detect the characteristic pion-decay feature at sub-GeV energies in the gamma-ray spectra. The two photons from the decay of the neutral pion ($\pi^0\rightarrow 2 \gamma$) have an isotropic decay spectrum in the rest frame of the neutral pion, leading to a characteristic spectral feature referred to as the pion-decay bump in the vicinity of 200~MeV \cite{1971NASSP.249.....S}. Using data from the {\it Fermi}-LAT, the identification of this so-called pion bump has been achieved in the spectra of the supernova remnants (SNRs) W44, IC~443 \cite{Fermi-LAT:2013iui}, W49B \cite{HESS:2016qan} and W51C  \cite{2016ApJ...816..100J}. More recently, a comprehensive search~\cite{Fermi-LAT:2022rkm} for such a feature among 4FGL sources within $5^\circ$ from the Galactic plane identified 56 sources with a significant spectral break. This population includes 13 SNRs, 6 sources of unknown nature but overlapping with known SNRs or PWNe, 4 binaries, 2 PWNe, 1 star-formation region, the Cygnus Cocoon, along with 30 unidentified sources of unknown origin. 

The second method is to use multiwavelength observations, such as the coincidence of the spatial locations of molecular clouds and gamma rays, the correlation in temporal profiles of gamma-ray and lower-energy emission of transients, and the broadband spectrum in general, to infer the origin of the gamma-ray emission. The following classes of gamma-ray sources have been the subject of intense study and have yielded promising candidate sources of cosmic-ray accelerators.

\begin{itemize}
  \item {\bf The Galactic Center}. The observation of gamma rays originating close to Sgr~A* with a photon index of $\sim 2.3$ extending to beyond 10~TeV suggests the existence of a PeVatron in the Galactic Center \cite{HESS:2016pst,HESS:2017tce,Adams:2021kwm,MAGIC:2020kxa}. The gamma-ray emission may be explained by protons accelerated by the black hole Sgr~A*, a nearby SNR, or the compact stellar clusters at the Galactic Center that are injected into the central molecular zone. The Galactic Center region has recently been detected by HAWC at yet higher energies \cite{Yun-Carcamo:2023Na}. 
  
  \item {\bf Star-formation regions and the Cygnus Cocoon}. Stellar clusters, composed of massive young stars, may plausibly accelerate protons to very high energy (VHE) through the stellar winds or the shocks of supernova remnants expanding inside the bubble \cite{Ackermann:2011lfa,Aharonian:2018oau,Cristofari:2021jkl}. Extended gamma-ray emission with a radius of more than $2^\circ$ was detected from the star-forming region of Cygnus~X by {\it Fermi}-LAT at 1-100~GeV, referred to as the Cygnus Cocoon \cite{Ackermann:2011lfa, Astiasarain:2023zal}. The gamma-ray flux cannot be explained by the Galactic diffuse emission, suggesting that either cosmic-ray electrons or protons are freshly accelerated inside the Cocoon. HAWC identified 1-100~TeV gamma rays from the Cygnus Cocoon, with a spatial morphology matching that observed at GeV energies remarkably well \cite{Abeysekara:2021yum}. The gamma-ray spectrum of the Cocoon softens from $dN/dE\propto E^{-2.2}$ at 1-100~GeV to $E^{-2.64}$ at 1-100~TeV, which may indicate the termination of the accelerator or the leakage of the highest energy particles. The matching morphology of the GeV and TeV emission, the spatial profile of the gamma rays, as well as the absence of diffuse X-ray emission from the Cocoon \cite{Mizuno:2015jua,Guevel:2022esw} all point to a hadronic source.
  
  The Cygnus region has also been observed by LHAASO \cite{LHAASO:2023uhj}. After removing all sources from the so-called Cygnus Bubble, which they define as a region of $6^\circ$ radius from the center, the residual emission can be fitted with a Gaussian template with a width of $2.28^\circ$ and $2.17^\circ$ using the WCDA and KM2A data, respectively. This flux of residual emission, named LHAASO~J2027$+$4119, and that from the entire Cygnus Bubble is compared to the {\it Fermi}-LAT and HAWC Cygnus Cocoon data in~\cref{fig:cocoon}.  In the extended vicinity of the Cygnus Cocoon, referred to by LHAASO as the Cygnus-X region, which is a $10^\circ$-radius extended region around Cygnus-X, photons with extreme energies up to 1.4~PeV have been observed~\cite{Tibet21, LHAASO:2023uhj}, hinting at the existence of a ``superPeVatron."
  
  In addition to the Cygnus Cocoon and the potential stellar cluster region at the Galactic Center, Westerlund 1 and Westerlund 2 are also detected at TeV energies \cite{2012A&A...537A.114A,2018A&A...611A..77Y}. More candidate gamma-ray emitting stellar clusters have been identified by comparing theoretical models to compact clusters found by Gaia \cite{Mitchell:2024yuy}. 
  
  \item {\bf Supernova remnants}. While SNRs represent the dominant population of GeV gamma-ray emitters with a pion bump feature, most of them either are undetected or have a steep spectrum at VHE of 0.1-100~TeV, challenging the theory that SNRs are the accelerators of cosmic rays up to the knee \cite{Aharonian:2018oau,Cristofari:2021jkl}. The scarcity of PeVatron candidates makes SNR~G106.3+2.7 a particularly interesting case. It is one of the nearest middle-aged SNRs, at $\sim800$~pc. Its emission above 10~TeV has been observed by the air shower gamma-ray detectors HAWC \cite{HAWC:2020nvc}, AS$\gamma$ \cite{TibetASg:2021kgt}, and LHAASO \cite{2021Natur.594...33C} and at multi-TeV energies by VERITAS \cite{Acciari:2009zz} and MAGIC \cite{MAGIC:2022iap}. The direction of the highest energy photons is spatially coincident with a molecular cloud \cite{TibetASg:2021kgt}. Also, the nondetection of 1-10~GeV counterparts to the VHE emission constrains the bremsstrahlung and inverse Compton emission by relativistic electrons, supporting a PeVatron scenario \cite{Fang:2022uge}. 
  
  \item {\bf Novae}. An outburst of RS Ophiuchi (RS Oph), a recurrent nova with a red giant companion, was observed by the LAT, MAGIC, and H.E.S.S. telescopes \cite{MAGIC:2022rmr, HESS:2022qap}. The similarity of the spectra and light curves at GeV and VHE energies favors a hadronic origin over the leptonic alternative.  The detection of time-correlated optical and gamma-ray emission in the classical novae V906 Carinae revealed the role of radiative shocks powering cosmic-ray acceleration \cite{Aydi:2020znu}. 
  These observations suggest that novae as well as other types of transients powered by non-relativistic shocks could be promising neutrino emitters \cite{Fang:2020bkm, 2021NatAs...5..472W, 2023PhRvD.108h3035G}. 
\end{itemize}

% challenges in identifying PeVatrons through photons alone (LHAASO UHE gamma sources; gamma-ray dim neutrino emitters) 
The detection of gamma-ray sources above 100~TeV, referred to as ultrahigh-energy (UHE) gamma-ray sources, provides opportunities to study both leptonic and hadronic gamma-ray emitters at the highest photon energies. To date, a total of 43 UHE gamma-ray sources have been detected at the $>4\sigma$ level \cite{HAWC:2019tcx, LHAASO:2023rpg}. A good fraction of these sources are found to be associated with pulsars, challenging the traditional paradigm that $>100$-TeV photons imply the presence of hadronic PeVatrons  \cite{2023AppSc..13.6433C}. 

The similarity in the spectra of pionic gamma rays and leptonic gamma rays from bremsstrahlung and inverse Compton scattering of relativistic electrons as well as the lack of multiwavelength counterparts are among the main challenges in identifying hadronic gamma-ray emitters. High-energy neutrinos, which predominantly originate from the decay of charged pions produced in hadronic interactions, provide a smoking gun for identifying the hadronic accelerators in the Milky Way. But this is also a challenging road, leaving the origin of the Galactic cosmic rays as one of the oldest puzzles in astronomy. We discuss this next.

\section{Neutrino Search for HAWC and LHAASO Sources}

Unlike the challenging but successful identification of the Milky Way, the search for neutrinos from the sources of the Galactic cosmic rays has come up empty so far. However, the multimessenger data suggest the presence of a component of unresolved sources in the Galactic neutrino flux observed by IceCube at the highest energies. This has led to a renewed effort to identify them.

% the Galactic plane observation: https://arxiv.org/abs/2307.04427
Building on a decade-long effort to identify the Galactic cosmic-ray sources, by matching gamma-ray and neutrino spectra in order to reveal their pionic origin~\cite{Alvarez-Muniz:2002con}, initiated
with the early IceCube data, a list of target Galactic sources has emerged  in IceCube's a priori defined source list, including 2HWC~J2031$+$415, Gamma Cygni, MGRO~J2019$+$37, MGRO~J1908$+$06, HESS~J1857$+$026, HESS~J1852$-$000, HESS~J1849$+$000, HESS~J1843$-$033, PSR~B0656$+$14, the Crab nebula, HESS~J1841$-$055, and HESS~J1837$-$069 \cite{IceCube:2019cia}. Additionally, source lists of supernova remnants, pulsar wind nebulae, and unidentified objects are used for IceCube's stacked catalogue searches \cite{IceCube:2019lzm,IceCube:2019cia,IceCube:2023ame}; see also Table S4 in the supplementary material of Ref.~\cite{IceCubeGP}. Some of the stacked Galactic source searches using the shower data sample that revealed the Galactic plane yield evidence exceeding the $>3\,\sigma$ level. However, this evidence cannot be separated from that for the diffuse emission template of the Galactic plane. 

In addition to these sources, several dedicated searches for Galactic sources have been carried out starting with the early MILAGRO sources. MGRO J1908$+$06 has been one of the prominent candidate sources \cite{Halzen:2016seh}. It has been included in IceCube’s triggered search for point-like and extended sources. A search for steady sources using 8 years of IceCube track data found this source as the second warmest source in the catalog, with a pre-trial p-value of 0.0088 \cite{IceCube:2018ndw}. It remains the most significant Galactic source in more recent searches \cite{IceCube:2019cia}. 

\begin{figure*}[t]
\centering
\includegraphics[width=0.79\textwidth]{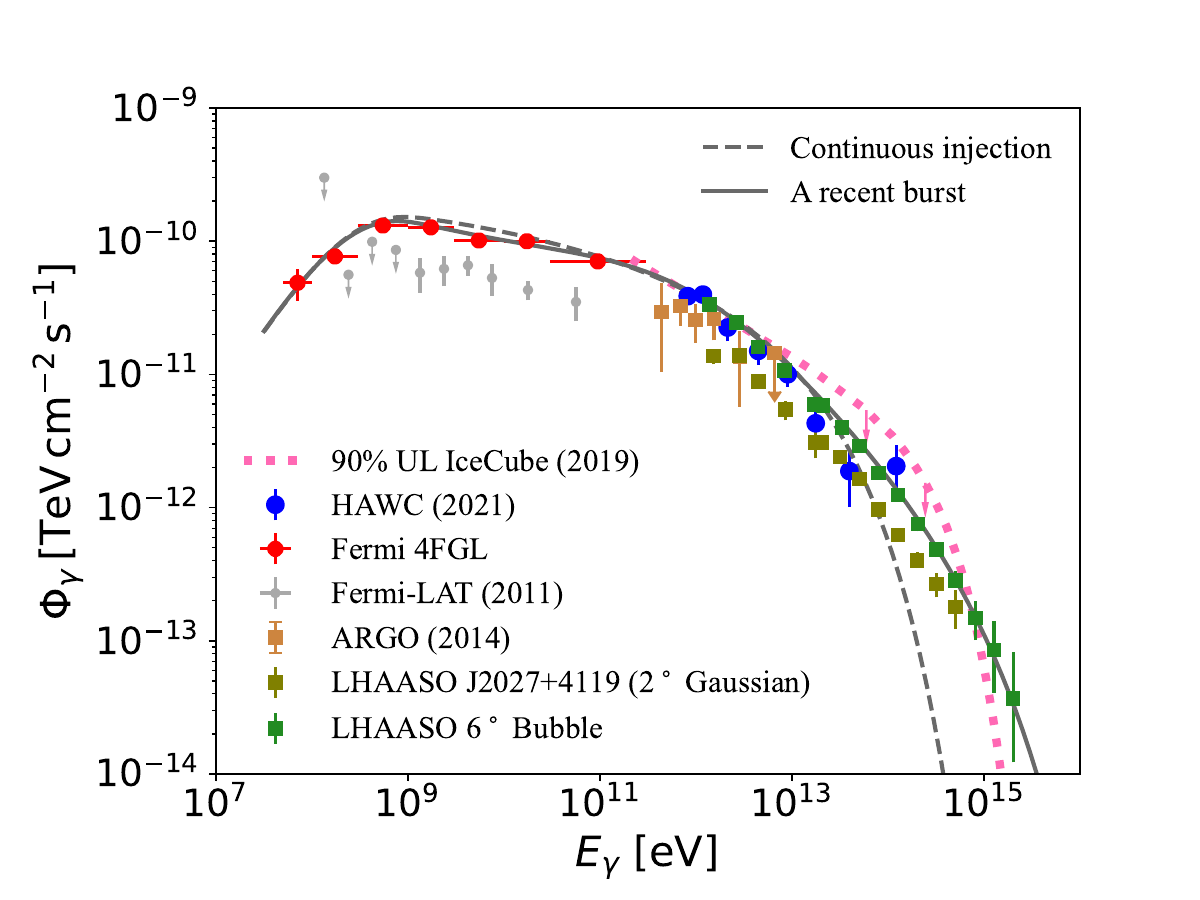}
 \caption{{\bf Spectral energy distribution of the gamma-ray emission at the Cocoon region}. The data points are from measurements by {\it Fermi}-LAT \cite{Ackermann:2011lfa, Abdollahi_2020}, ARGO \cite{ARGO-YBJ:2014ikx}, HAWC \cite{Abeysekara:2021yum}, and LHAASO \cite{LHAASO:2023uhj}. See Section~5 for more discussion regarding the gamma-ray observations of the Cygnus Cocoon. The solid and dashed curves denote two theory models from \cite{Abeysekara:2021yum}, which invoke continuous proton injection and a recent burst by a PeVatron, respectively.  The magenta dotted curve indicates an upper limit converted from IceCube's neutrino observation of the Cygnus region \cite{Kheirandish:2019bke}.    } 
\label{fig:cocoon} 
\end{figure*}

The IceCube and HAWC collaborations performed a joint analysis \cite{Kheirandish:2019bke} to examine the correlation of neutrinos and 2HWC sources \cite{Abeysekara:2017hyn}. Upper limits on various Galactic sources were derived. A comparison of neutrino and gamma-ray observations of the Cygnus Cocoon is shown in~\cref{fig:cocoon}. The joint analysis has been updated using more recent IceCube data and expanded HAWC catalogues.  

A variety of stacking analysis were performed searching for the neutrino emission from classes of strong gamma-ray emitters. Using 9.5 years of all-sky data, IceCube performed a stacking search for the neutrino emission from 35 PWNe with gamma-ray emission above 1~TeV assuming a variety of weighting schemes for the sources. The analysis constrained the hadronic component of the gamma-ray emission above $\sim 10$~TeV assuming that the neutrino flux of PWNe scales as the gamma-ray flux or as the inverse of the pulsar age~\cite{IceCube:2020svz}.

IceCube searched for neutrino emission from Galactic X-ray binaries in 7.5 years of data~\cite{IceCube:2022jpz}. In particular, they searched for a periodic neutrino emission from 55 binaries in the northern sky as well as for a time-dependent emission using the X-ray light curves of 102 binaries in the entire sky. Finally, a search was performed for the time-integrated emission of neutrinos from four microquasars: Cyg~X-3, LS~5039, LS~I~61$+$303, and SS~433. No significant signal was found in the periodic and flare analysis, while in the latter, the most significant excess was found for Cyg~X-3 with a post-trial p-value of 0.036, corresponding to only $1.8\,\sigma$.

% limits on LHAASO’s 12 UHE sources https://arxiv.org/abs/2211.14184 
With the advent of LHAASO data, IceCube performed a search for neutrinos from the 12 UHE gamma-ray sources observed by LHAASO based on 11 years of muon track events~\cite{IceCube:2022heu}. The analysis assumes an extension of the sources based on the gamma-ray observations. The neutrino spectrum is assumed to follow a power law with index between 1 and 4. Only upper limits on the hadronic component of the sources emerged. IceCube places constraints on the fraction of the gamma-ray flux originating from hadronic processes only for the Crab Nebula and LHAASO J2226+6057 at 59\% and 85\% of the total, when adopting a power-law spectrum as reported by LHAASO and a log-parabola spectrum by HAWC, respectively. In Ref.~\cite{IceCube:2022heu}, IceCube also performed stacking searches for the LHAASO UHE sources as well as for sources lists associated with PWNe and SNRs.

Recently, IceCube and HAWC performed a joint search for a correlation between the third HAWC gamma-ray catalog and IceCube point-source data \cite{Alfaro:2024suy}. No significant neutrino excess was identified from 22 gamma-ray sources. Among them, the upper limits on five sources, namely 3HWC~J1847$-$017, 3HWC~J1914$+$118, 3HWC~J1922$+$140, 3HWC~J0534$+$220, and 3HWC~J2227$+$610, are below the expected neutrino flux had the gamma-ray emission been 100\% pionic. The analysis limits the fraction of the hadronic component of these sources to $\sim$50-80\% level.

% searches for extended sources in the Galactic plane: https://arxiv.org/abs/2307.07576
Finally, using ten years of track-like events, IceCube searched for extended neutrino sources in the Galaxy~\cite{IceCube:2023ujd}. They searched the neutrino sky for sources with possible extensions of 0.5, 1.0, 1.5, or 2.0 degrees. No evidence for time-integrated neutrino emission from potentially extended sources was found. In a search for the neutrino emission from 20 regions of interest coincident with TeV gamma-ray sources, the most significant region was found to be centered at 3HWC~J1951$+$266, with a global significance of $2.6\,\sigma$ for an extension of $1.7^\circ$.

%\section{The Cygnus Cocoon/Bubble}
\section{Extragalactic Sources}

%https://arxiv.org/abs/2302.05459
Above 10~TeV, most sources are local as a result of the absorption of gamma rays by the extragalactic background light, though a handful of extragalactic sources have been observed \cite{HAWC:2020hrt, LHAASO:2023rpg}. Among them the gamma-ray burst GRB~ 221009A, the brightest GRB observed by Fermi-GBM to date, stands out as the first GRB observed above 10~TeV \cite{LHAASO:2023lkv}.  LHAASO's WCDA observed the first main pulse of GRB~221009A at 225~s to 228~s after the GBM trigger \cite{LHAASO:2023kyg}. The peak times for different energy bands agree with each other, suggesting that the peak corresponds to the onset of the afterglow. The TeV emission of GRB 221009A can be understood as synchrotron self-Compton (SSC) emission of relativistic electrons in the external shock, which is consistent with the interpretation of previous TeV afterglows \cite{MAGIC:2019irs,Abdalla:2019dlr}. No gamma ray from the prompt phase was observed, likely due to the high optical depth of pair production at the early times. 

Neutrino telescopes observed GRB~221009A with both real-time and follow-up searches over a wide range of energies \cite{IceCube:2023rhf, Aiello:2024eyp}. No significant deviation from background was found from MeV to PeV energies. The nondetection places tight constraints on models for neutrino emission during the prompt phase of GRBs. 

In the prompt phase, neutrinos can be produced by both nonthermal and thermal protons \cite{Meszaros:2015krr,Kimura:2022zyg}. Baryons entrained in the outflow can be accelerated in internal shocks, or by magnetic reconnection in Poynting-flux-dominated outflows. The accelerated protons may interact with background photons and produce TeV–PeV neutrinos in the internal shocks \cite{PhysRevLett.78.2292,2012PhRvL.108w1101H,PhysRevLett.110.121101,Bustamante:2014oka}. The nondetection of neutrinos from the single GRB~221009A constrains the nonthermal baryon-loading factor in the internal shock model, defined as the ratio of the fireball energy carried by protons and photons \cite{PhysRevD.73.063002} $\xi_p = E_{p,\rm iso}/E_{\gamma,\rm iso}$, with a sensitivity that matches the constraints obtained by stacking 1172 GRBs over the history of IceCube observations~\cite{IceCube:2017amx, Murase:2022vqf}.

Neutrinos may also be produced without cosmic-ray acceleration. ``Quasithermal” neutrinos may be produced by internal collisions between differentially streaming protons and neutrons as a result of the decoupling of the electrically neutral neutrons \cite{Bahcall:2000sa} (decoupling scenario) or of the transverse differences of the bulk Lorentz factor of the outflows  \cite{2010MNRAS.407.1033B, Murase:2013hh,  PhysRevLett.110.241101, PhysRevD.105.083023} (collision scenario). Assuming a Lorentz factor $\Gamma\sim 440$ \cite{LHAASO:2023kyg} and an isotropic-equivalent energy release by the burst $E_{\gamma,\rm iso} \sim 3\times 10^{54}\,\rm erg$ \cite{Burns:2023oxn}, the nondetection of quasithermal neutrinos in the IceCube low-energy DeepCore sample (GRECO sample) \cite{IceCube:2023rhf} constrains the baryon-loading factor to $\xi_p\sim 0.5$ in the collision scenario. 

PeV- to EeV-energy neutrino emission, which is predicted to result from ultrahigh-energy cosmic rays accelerated by the external shocks during the afterglow phase  \cite{Waxman:1999ai,Dermer:2000yd, PhysRevD.73.063002}, is not well constrained. Better sensitivities at these energies provided by future telescopes such as IceCube-Gen2 \cite{IceCube-Gen2:2020qha} will enable either observing or constraining the afterglow neutrino emission.

\section{Summary}
We reviewed the recent astrophysical observations by the IceCube Neutrino Observatory. We discussed the observations of the diffuse cosmic neutrino flux and the Galactic plane from the perspective of multimessenger observations. 

Inside the Milky Way, gamma-ray observations with AS$\gamma$, HAWC, and LHAASO reveal a bright Galactic diffuse background at tens to hundreds of TeV energies as well as a large, diverse population of very high and ultra-high-energy gamma-ray sources. The flux of neutrinos observed by IceCube from the Galactic plane is consistent with the gamma-ray Galactic diffuse emission measurements, though it may also be composed of individual pionic gamma-ray sources. High-energy neutrinos provide a clean venue for identifying the Galactic hadronic PeVatrons, though we also emphasized the importance of multimessenger approaches that combine the neutrino and gamma-ray observations. 

In the extragalactic sky, observations of the diffuse neutrino background and the first neutrino sources hint at the production of the bulk of high-energy neutrinos in gamma-ray-opaque environments. A comparison of the Galactic and extragalactic neutrino fluxes suggests that a distant Milky Way-like galaxy could be orders of magnitude more luminous in neutrinos than our own Galaxy. 

In general, it is clear that a next generation of neutrino telescopes is required to move from discovery to astronomy. In particular, we anticipate the identification of the cosmic-ray sources in the Galaxy using the multimessenger approach described in this review.

\section{Acknowledgements}
The work of K.F and F.H is supported by the Office of the Vice Chancellor for Research at the University of Wisconsin--Madison with funding from the Wisconsin Alumni Research Foundation. K.F. acknowledges support from the National Science Foundation (PHY-2110821, PHY-2238916) and NASA (NMH211ZDA001N-{\it Fermi}). This work was supported by a grant from the Simons Foundation (00001470, KF). K.F acknowledges the support of the Sloan Research Fellowship. The research of F.H was also supported in part by the U.S. National Science Foundation under grants~PHY-2209445 and OPP-2042807.

% =================================
% REFERENCES
% =================================
%%%%%%%%%%%%%%%%%%%%%%%%%%%%%%%%%%%%%%%%%%%%%%%%%%%%%%%%%%%%%%%%%
\newcommand{\astropart} {Astropart. Phys.}
\newcommand {\araa}     {Annual Review of Astronomy and Astrophysics}
\newcommand {\actaa}    {Acta Astronomica}
\newcommand {\ao}       {Applied Optics}
\newcommand {\aj}       {Astronomical Journal}
\newcommand {\azh}      {Astronomicheskii Zhurnal}
\newcommand {\aap}      {A\&A}
\newcommand {\aapr}     {A\&A Reviews}
\newcommand {\apj}      {ApJ.}
\newcommand {\apjl}     {ApJ Letters}
\newcommand {\apjs}     {The Astrophysical Journal Supplement Series}
\newcommand {\aplett}   {Astrophysics Letters and Communications}
\newcommand {\apspr}    {Astrophysics Space Physics Research}
\newcommand {\apss}     {Astrophysics and Space Science}
\newcommand {\aaps}     {Astrophysics and Space Science Supplement Series}
\newcommand {\brjp}     {Brazilian Journal of Physics}
\newcommand {\baas}     {Bulletin of the AAS}
\newcommand {\bain}     {Bulletin Astronomical Inst. Netherlands}
\newcommand {\caa}      {Chinese Astronomy and Astrophysics}
\newcommand {\cjaa}     {Chinese Journal of Astronomy and Astrophysics}
\newcommand {\coasp}    {Comments on Astrophysics and Space Physics}
\newcommand {\fcp}      {Fundamental Cosmic Physics}
\newcommand {\grg}      {General Relativity and Gravitation}
\newcommand {\gca}      {Geochimica Cosmochimica Acta}
\newcommand {\grl}      {Geophysics Research Letters}
\newcommand {\iaucirc}  {IAU Circular}
\newcommand {\icarus}   {Icarus}
\newcommand {\ijmpa}    {International Journal of Modern Physics A}
\newcommand {\ijmpb}    {International Journal of Modern Physics B}
\newcommand {\ijmpc}    {International Journal of Modern Physics C}
\newcommand {\ijmpd}    {International Journal of Modern Physics D}
\newcommand {\ijmpe}    {International Journal of Modern Physics E}
\newcommand {\jcp}      {Journal of Chemical Physics}
\newcommand {\jgr}      {Journal of Geophysical Research}
\newcommand {\jcap}     {JCAP}
\newcommand {\jqsrt}    {Journal of Quantitative Specstroscopy and Radiative Transfer}
\newcommand {\jrasc}    {Journal of the RAS of Canada}
\newcommand {\memras}   {Memoirs of the RAS}
\newcommand {\memsai}   {Mem. Societa Astronomica Italiana}
\newcommand {\mnras}    {MNRAS}
\newcommand {\na}       {New Astronomy}
\newcommand {\nar}      {New Astronomy Review}
\newcommand {\nat}      {Nature}
\newcommand {\nphysa}   {Nuclear Physics A}
\newcommand {\nphysb}   {Nuclear Physics B}
\newcommand {\nphysbs}  {Nuclear Physics B Proceedings Supplements}
\newcommand {\physscr}  {Physica Scripta}
\newcommand {\pr}       {Physical Review}
\newcommand {\pra}      {Physical Review A}
\newcommand {\prb}      {Physical Review B}
\newcommand {\prc}      {Physical Review C}
\newcommand {\prd}      {Physical Review D}
\newcommand {\pre}      {Physical Review E}
\newcommand {\PRL}      {Phys. Rev. Lett.}
\newcommand {\prl}      {Phys. Rev. Lett.}
\newcommand {\physrep}  {Physics Reports}
\newcommand {\planss}   {Planetary Space Science}
\newcommand {\pnas}     {Proceedings of the National Academy of Science}
\newcommand {\procspie} {Proceedings of the SPIE}
\newcommand {\pasa}     {Publications of the ASA}
\newcommand {\pasj}     {Publications of the ASJ}
\newcommand {\pasp}     {Publications of the ASP}
\newcommand {\qjras}    {Quarterly Journal of the RAS}
\newcommand {\rmp}      {Reviews of Modern Physics}
\newcommand {\ssr}      {Space Science Reviews}
\newcommand\mathplus{+}

\bibliographystyle{elsarticle-num}
\bibliography{sample}

@article{Alvarez-Muniz:2002con,
    author = "Alvarez-Muniz, J. and Halzen, F.",
    title = "{Possible high-energy neutrinos from the cosmic accelerator RX J1713.7-3946}",
    eprint = "astro-ph/0205408",
    archivePrefix = "arXiv",
    reportNumber = "MADPH-02-1272",
    doi = "10.1086/342978",
    journal = "Astrophys. J. Lett.",
    volume = "576",
    pages = "L33--L36",
    year = "2002"
}

@article{Kato:2024kxn,
    author = "Kato, S. and others",
    title = "{On the source contribution to the Galactic diffuse gamma rays above 398 TeV detected by the Tibet AS\ensuremath{\gamma} experiment}",
    eprint = "2401.13120",
    archivePrefix = "arXiv",
    primaryClass = "astro-ph.HE",
    month = "1",
    year = "2024"
}

@ARTICLE{2022ApJ...928...19V,
       author = {{Vecchiotti}, V. and {Zuccarini}, F. and {Villante}, F.~L. and {Pagliaroli}, G.},
        title = "{Unresolved Sources Naturally Contribute to PeV Gamma-Ray Diffuse Emission Observed by Tibet AS{\ensuremath{\gamma}}}",
      journal = {\apj},
         year = 2022,
        month = mar,
       volume = {928},
       number = {1},
          eid = {19},
        pages = {19},
          doi = {10.3847/1538-4357/ac4df4},
archivePrefix = {arXiv},
       eprint = {2107.14584},
 primaryClass = {astro-ph.HE},
       adsurl = {https://ui.adsabs.harvard.edu/abs/2022ApJ...928...19V}
}

@ARTICLE{2016A&A...595A..58C,
       author = {{Cardillo}, M. and {Amato}, E. and {Blasi}, P.},
        title = "{Supernova remnant W44: a case of cosmic-ray reacceleration}",
      journal = {\aap},
         year = 2016,
        month = oct,
       volume = {595},
          eid = {A58},
        pages = {A58},
          doi = {10.1051/0004-6361/201628669},
archivePrefix = {arXiv},
       eprint = {1604.02321},
 primaryClass = {astro-ph.HE},
       adsurl = {https://ui.adsabs.harvard.edu/abs/2016A&A...595A..58C}
}

@article{Mitchell:2024yuy,
    author = "Mitchell, Alison M. W. and Morlino, Giovanni and Celli, Silvia and Menchiari, Stefano and Specovius, Andreas",
    title = "{Probing Stellar Clusters from Gaia DR2 as Galactic PeVatrons: I -- Expected Gamma-ray and Neutrino Emission}",
    eprint = "2403.16650",
    archivePrefix = "arXiv",
    primaryClass = "astro-ph.HE",
    month = "3",
    year = "2024"
}

@ARTICLE{2012A&A...537A.114A,
       author = {{Abramowski}, A. and {Acero}, F. and {Aharonian}, F. and {Akhperjanian}, A.~G. and {Anton}, G. and others},
        title = "{Discovery of extended VHE {\ensuremath{\gamma}}-ray emission from the vicinity of the young massive stellar cluster Westerlund 1}",
      journal = {\aap},
         year = 2012,
        month = jan,
       volume = {537},
          eid = {A114},
        pages = {A114},
          doi = {10.1051/0004-6361/201117928},
archivePrefix = {arXiv},
       eprint = {1111.2043},
 primaryClass = {astro-ph.HE},
       adsurl = {https://ui.adsabs.harvard.edu/abs/2012A&A...537A.114A}
}

@ARTICLE{2018A&A...611A..77Y,
       author = {{Yang}, Rui-zhi and {de O{\~n}a Wilhelmi}, Emma and {Aharonian}, Felix},
        title = "{Diffuse {\ensuremath{\gamma}}-ray emission in the vicinity of young star cluster Westerlund 2}",
      journal = {\aap},
         year = 2018,
        month = apr,
       volume = {611},
          eid = {A77},
        pages = {A77},
          doi = {10.1051/0004-6361/201732045},
archivePrefix = {arXiv},
       eprint = {1710.02803},
 primaryClass = {astro-ph.HE},
       adsurl = {https://ui.adsabs.harvard.edu/abs/2018A&A...611A..77Y}
}

@article{LHAASO:2023uhj,
    author = "Cao, Zhen and others",
    collaboration = "LHAASO",
    title = "{An ultrahigh-energy \ensuremath{\gamma}-ray bubble powered by a super PeVatron}",
    eprint = "2310.10100",
    archivePrefix = "arXiv",
    primaryClass = "astro-ph.HE",
    doi = "10.1016/j.scib.2023.12.040",
    journal = "Sci. Bull.",
    volume = "69",
    number = "4",
    pages = "449--457",
    year = "2024"
}

@article{Mizuno:2015jua,
    author = "Mizuno, T. and Tanabe, T. and Takahashi, H. and Hayashi, K. and Yamazaki, R. and Grenier, I. and Tibaldo, L.",
    title = "{$Suzaku$ Observation of the Fermi Cygnus Cocoon: the Search for a Signature of Young Cosmic-ray Electrons}",
    eprint = "1502.01390",
    archivePrefix = "arXiv",
    primaryClass = "astro-ph.HE",
    doi = "10.1088/0004-637X/803/2/74",
    journal = "Astrophys. J.",
    volume = "803",
    number = "2",
    pages = "74",
    year = "2015"
}

@article{Astiasarain:2023zal,
    author = {Astiasarain, X. and Tibaldo., L. and Martin, P. and Kn\"odlseder, J. and Remy, Q.},
    title = "{Multiple emission components in the Cygnus cocoon detected from Fermi-LAT observations}",
    eprint = "2301.04504",
    archivePrefix = "arXiv",
    primaryClass = "astro-ph.HE",
    doi = "10.1051/0004-6361/202245573",
    journal = "Astron. Astrophys.",
    volume = "671",
    pages = "A47",
    year = "2023"
}

@article{Guevel:2022esw,
    author = "Guevel, David and Beardmore, Andrew and Page, Kim L. and Lien, Amy and Fang, Ke and Tibaldo, Luigi and Casanova, Sabrina and Huentemeyer, Petra",
    title = "{Limits on Leptonic TeV Emission from the Cygnus Cocoon with Swift-XRT}",
    eprint = "2211.07617",
    archivePrefix = "arXiv",
    primaryClass = "astro-ph.HE",
    doi = "10.3847/1538-4357/accdde",
    journal = "Astrophys. J.",
    volume = "950",
    number = "2",
    pages = "116",
    year = "2023"
}

@article{Abeysekara:2021yum,
    author = "Abeysekara, A. U. and others",
    title = "{Author Correction: HAWC observations of the acceleration of very-high-energy cosmic rays in the Cygnus Cocoon [doi: 10.1038/s41550-021-01318-y]}",
    eprint = "2103.06820",
    archivePrefix = "arXiv",
    primaryClass = "astro-ph.HE",
    doi = "10.1038/s41550-021-01361-9",
    journal = "Nature Astron.",
    volume = "5",
    number = "5",
    pages = "465--471",
    year = "2021",
    note = "[Erratum: Nature Astron. 5, 724--724 (2021)]"
}

@article{Ackermann:2011lfa,
    author = "Ackermann, M. and others",
    title = "{A Cocoon of Freshly Accelerated Cosmic Rays Detected by Fermi in the Cygnus Superbubble}",
    doi = "10.1126/science.1210311",
    journal = "Science",
    volume = "334",
    number = "6059",
    pages = "1103--1107",
    year = "2011"
}

@article{MAGIC:2022iap,
    author = "Abe, H. and others",
    collaboration = "MAGIC",
    title = "{MAGIC observations provide compelling evidence of hadronic multi-TeV emission from the putative PeVatron SNR G106.3+2.7}",
    eprint = "2211.15321",
    archivePrefix = "arXiv",
    primaryClass = "astro-ph.HE",
    doi = "10.1051/0004-6361/202244931",
    journal = "Astron. Astrophys.",
    volume = "671",
    pages = "A12",
    year = "2023"
}

@article{TibetASg:2021kgt,
    author = "Amenomori, M. and others",
    collaboration = "Tibet AS\ensuremath{\gamma}",
    title = "{Potential PeVatron supernova remnant G106.3+2.7 seen in the highest-energy gamma rays}",
    eprint = "2109.02898",
    archivePrefix = "arXiv",
    primaryClass = "astro-ph.HE",
    doi = "10.1038/s41550-020-01294-9",
    journal = "Nature Astron.",
    volume = "5",
    number = "5",
    pages = "460--464",
    year = "2021"
}

@ARTICLE{2021Natur.594...33C,
       author = {{Cao}, Zhen and {Aharonian}, F.~A. and {An}, Q. and {Axikegu}, Bai, L.~X. and {Bai}, Y.~X. and others},
        title = "{Ultrahigh-energy photons up to 1.4 petaelectronvolts from 12 {\ensuremath{\gamma}}-ray Galactic sources}",
      journal = {\nat},
         year = 2021,
        month = jun,
       volume = {594},
       number = {7861},
        pages = {33-36},
          doi = {10.1038/s41586-021-03498-z},
       adsurl = {https://ui.adsabs.harvard.edu/abs/2021Natur.594...33C}
}

@article{HAWC:2020nvc,
    author = "Albert, A. and others",
    collaboration = "HAWC",
    title = "{HAWC J2227+610 and its association with G106.3+2.7, a new potential Galactic PeVatron}",
    eprint = "2005.13699",
    archivePrefix = "arXiv",
    primaryClass = "astro-ph.HE",
    doi = "10.3847/2041-8213/ab96cc",
    journal = "Astrophys. J. Lett.",
    volume = "896",
    pages = "L29",
    year = "2020"
}

@article{Aharonian:2018oau,
    author = "Aharonian, Felix and Yang, Ruizhi and de O\~na Wilhelmi, Emma",
    title = "{Massive Stars as Major Factories of Galactic Cosmic Rays}",
    eprint = "1804.02331",
    archivePrefix = "arXiv",
    primaryClass = "astro-ph.HE",
    doi = "10.1038/s41550-019-0724-0",
    journal = "Nature Astron.",
    volume = "3",
    number = "6",
    pages = "561--567",
    year = "2019"
}

@article{Cristofari:2021jkl,
    author = "Cristofari, Pierre",
    title = "{The Hunt for Pevatrons: The Case of Supernova Remnants}",
    eprint = "2110.07956",
    archivePrefix = "arXiv",
    primaryClass = "astro-ph.HE",
    doi = "10.3390/universe7090324",
    journal = "Universe",
    volume = "7",
    number = "9",
    pages = "324",
    year = "2021"
}

@article{Fang:2022uge,
    author = "Fang, Ke and Kerr, Matthew and Blandford, Roger and Fleischhack, Henrike and Charles, Eric",
    title = "{Evidence for PeV Proton Acceleration from Fermi-LAT Observations of SNR G106.3+2.7}",
    eprint = "2208.05457",
    archivePrefix = "arXiv",
    primaryClass = "astro-ph.HE",
    doi = "10.1103/PhysRevLett.129.071101",
    journal = "Phys. Rev. Lett.",
    volume = "129",
    number = "7",
    pages = "071101",
    year = "2022"
}

@inproceedings{Yun-Carcamo:2023Na,
  author = "Yun-Carcamo, Sohyoun L.  and  Albert, Andrea  and  Alfaro, Ruben Jose  and  Alvarez, César  and  Andres, Alexis  and others",
  title = "{Observation of the Crab Nebula and Galactic Center with the improved HAWC reconstruction algorithm}",
  doi = "10.22323/1.444.0697",
  booktitle = "Proceedings of 38th International Cosmic Ray Conference {\textemdash} PoS(ICRC2023)",
  year = 2023,
  volume = "444",
  pages = "697"
}

@article{MAGIC:2020kxa,
    author = "Acciari, V. A. and others",
    collaboration = "MAGIC",
    title = "{MAGIC observations of the diffuse \ensuremath{\gamma}-ray emission in the vicinity of the Galactic center}",
    eprint = "2006.00623",
    archivePrefix = "arXiv",
    primaryClass = "astro-ph.HE",
    doi = "10.1051/0004-6361/201936896",
    journal = "Astron. Astrophys.",
    volume = "642",
    pages = "A190",
    year = "2020"
}

@article{Adams:2021kwm,
    author = "Adams, C. B. and others",
    title = "{VERITAS Observations of the Galactic Center Region at Multi-TeV Gamma-Ray Energies}",
    eprint = "2104.12735",
    archivePrefix = "arXiv",
    primaryClass = "astro-ph.HE",
    doi = "10.3847/1538-4357/abf926",
    journal = "Astrophys. J.",
    volume = "913",
    number = "2",
    pages = "115",
    year = "2021"
}

@article{HESS:2017tce,
    author = "Abdalla, H. and others",
    collaboration = "HESS",
    title = "{Characterising the VHE diffuse emission in the central 200 parsecs of our Galaxy with H.E.S.S}",
    eprint = "1706.04535",
    archivePrefix = "arXiv",
    primaryClass = "astro-ph.HE",
    doi = "10.1051/0004-6361/201730824",
    journal = "Astron. Astrophys.",
    volume = "612",
    pages = "A9",
    year = "2018"
}

@article{HESS:2016pst,
    author = "Abramowski, A. and others",
    collaboration = "H.E.S.S.",
    title = "{Acceleration of petaelectronvolt protons in the Galactic Centre}",
    eprint = "1603.07730",
    archivePrefix = "arXiv",
    primaryClass = "astro-ph.HE",
    doi = "10.1038/nature17147",
    journal = "Nature",
    volume = "531",
    pages = "476",
    year = "2016"
}

@article{Aydi:2020znu,
    author = "Aydi, Elias and others",
    title = "{Direct evidence for shock-powered optical emission in a nova}",
    eprint = "2004.05562",
    archivePrefix = "arXiv",
    primaryClass = "astro-ph.HE",
    doi = "10.1038/s41550-020-1070-y",
    journal = "Nature Astron.",
    volume = "4",
    number = "8",
    pages = "776--780",
    year = "2020"
}

@ARTICLE{2021NatAs...5..472W,
       author = {{Winter}, Walter and {Lunardini}, Cecilia},
        title = "{A concordance scenario for the observed neutrino from a tidal disruption event}",
      journal = {Nature Astronomy},
         year = 2021,
        month = may,
       volume = {5},
        pages = {472-477},
          doi = {10.1038/s41550-021-01305-3},
archivePrefix = {arXiv},
       eprint = {2005.06097},
 primaryClass = {astro-ph.HE},
       adsurl = {https://ui.adsabs.harvard.edu/abs/2021NatAs...5..472W}
}

@ARTICLE{2023PhRvD.108h3035G,
       author = {{Guarini}, Ersilia and {Tamborra}, Irene and {Margutti}, Raffaella and {Ramirez-Ruiz}, Enrico},
        title = "{Transients stemming from collapsing massive stars: The missing pieces to advance joint observations of photons and high-energy neutrinos}",
      journal = {\prd},
         year = 2023,
        month = oct,
       volume = {108},
       number = {8},
          eid = {083035},
        pages = {083035},
          doi = {10.1103/PhysRevD.108.083035},
archivePrefix = {arXiv},
       eprint = {2308.03840},
 primaryClass = {astro-ph.HE},
       adsurl = {https://ui.adsabs.harvard.edu/abs/2023PhRvD.108h3035G}
}

@article{Fang:2020bkm,
    author = "Fang, Ke and Metzger, Brian D. and Vurm, Indrek and Aydi, Elias and Chomiuk, Laura",
    title = "{High-energy Neutrinos and Gamma Rays from Nonrelativistic Shock-powered Transients}",
    eprint = "2007.15742",
    archivePrefix = "arXiv",
    primaryClass = "astro-ph.HE",
    doi = "10.3847/1538-4357/abbc6e",
    journal = "Astrophys. J.",
    volume = "904",
    number = "1",
    pages = "4",
    year = "2020"
}

@article{HESS:2022qap,
    author = "De Wolf, Stefaan and others",
    collaboration = "H.E.S.S.",
    title = "{Time-resolved hadronic particle acceleration in the recurrent nova RS~Ophiuchi}",
    eprint = "2202.08201",
    archivePrefix = "arXiv",
    primaryClass = "astro-ph.HE",
    doi = "10.1126/science.abn0567",
    journal = "Science",
    volume = "376",
    number = "6588",
    pages = "abn0567",
    year = "2022"
}

@article{MAGIC:2022rmr,
    author = "Acciari, V. A. and others",
    collaboration = "MAGIC",
    title = "{Proton acceleration in thermonuclear nova explosions revealed by gamma rays}",
    eprint = "2202.07681",
    archivePrefix = "arXiv",
    primaryClass = "astro-ph.HE",
    doi = "10.1038/s41550-022-01640-z",
    journal = "Nature Astron.",
    volume = "6",
    pages = "689--697",
    year = "2022",
    note = "[Erratum: Nature Astron. 6, 760--760 (2022)]"
}

@article{Fermi-LAT:2022rkm,
    author = "Abdollahi, S. and others",
    collaboration = "Fermi-LAT",
    title = "{Search for New Cosmic-Ray Acceleration Sites within the 4FGL Catalog Galactic Plane Sources}",
    eprint = "2205.03111",
    archivePrefix = "arXiv",
    primaryClass = "astro-ph.HE",
    doi = "10.3847/1538-4357/ac704f",
    journal = "Astrophys. J.",
    volume = "933",
    number = "2",
    pages = "204",
    year = "2022"
}

@ARTICLE{2010ApJ...723L.122U,
       author = {{Uchiyama}, Yasunobu and {Blandford}, Roger D. and {Funk}, Stefan and {Tajima}, Hiroyasu and {Tanaka}, Takaaki},
        title = "{Gamma-ray Emission from Crushed Clouds in Supernova Remnants}",
      journal = {\apjl},
         year = 2010,
        month = nov,
       volume = {723},
       number = {1},
        pages = {L122-L126},
          doi = {10.1088/2041-8205/723/1/L122},
archivePrefix = {arXiv},
       eprint = {1008.1840},
 primaryClass = {astro-ph.HE},
       adsurl = {https://ui.adsabs.harvard.edu/abs/2010ApJ...723L.122U}
}

@article{HESS:2016qan,
    author = "Abdalla, H. and others",
    collaboration = "H.E.S.S., Fermi-LAT",
    title = "{The supernova remnant W49B as seen with H.E.S.S. and Fermi-LAT}",
    eprint = "1609.00600",
    archivePrefix = "arXiv",
    primaryClass = "astro-ph.HE",
    doi = "10.1051/0004-6361/201527843",
    journal = "Astron. Astrophys.",
    volume = "612",
    pages = "A5",
    year = "2018"
}

@ARTICLE{2016ApJ...816..100J,
       author = {{Jogler}, T. and {Funk}, S.},
        title = "{Revealing W51C as a Cosmic Ray Source Using Fermi-LAT Data}",
      journal = {\apj},
         year = 2016,
        month = jan,
       volume = {816},
       number = {2},
          eid = {100},
        pages = {100},
          doi = {10.3847/0004-637X/816/2/100},
       adsurl = {https://ui.adsabs.harvard.edu/abs/2016ApJ...816..100J}
}

@book{1971NASSP.249.....S,
  title = {Cosmic Gamma Rays},
  author = {Stecker, F. W.},
  year = {1971},
  month = jan,
  publisher = {U.S. Govt. Printing Office, NASA-SP249,19710015288.pdf}, 
  
}

@article{Fermi-LAT:2013iui,
    author = "Ackermann, M. and others",
    collaboration = "Fermi-LAT",
    title = "{Detection of the Characteristic Pion-Decay Signature in Supernova Remnants}",
    eprint = "1302.3307",
    archivePrefix = "arXiv",
    primaryClass = "astro-ph.HE",
    doi = "10.1126/science.1231160",
    journal = "Science",
    volume = "339",
    pages = "807",
    year = "2013"
}

@ARTICLE{2015ApJ...815L..25G,
       author = {{Gaggero}, Daniele and {Grasso}, Dario and {Marinelli}, Antonio and {Urbano}, Alfredo and {Valli}, Mauro},
        title = "{The Gamma-Ray and Neutrino Sky: A Consistent Picture of Fermi-LAT, Milagro, and IceCube Results}",
      journal = {\apjl},
         year = 2015,
        month = dec,
       volume = {815},
       number = {2},
          eid = {L25},
        pages = {L25},
          doi = {10.1088/2041-8205/815/2/L25},
archivePrefix = {arXiv},
       eprint = {1504.00227},
 primaryClass = {astro-ph.HE},
       adsurl = {https://ui.adsabs.harvard.edu/abs/2015ApJ...815L..25G}
}

@ARTICLE{lhaasoGDE,
       author = {{Cao}, Zhen and {Aharonian}, F. and {An}, Q. and {Axikegu} and {Bai}, Y.~X. and others},
        title = "{Measurement of ultra-high-energy diffuse gamma-ray emission of the Galactic plane from 10 TeV to 1 PeV with LHAASO-KM2A}",
      journal = {arXiv e-prints},
         year = 2023,
        month = may,
          eid = {arXiv:2305.05372},
        pages = {arXiv:2305.05372},
          doi = {10.48550/arXiv.2305.05372},
archivePrefix = {arXiv},
       eprint = {2305.05372},
 primaryClass = {astro-ph.HE},
       adsurl = {https://ui.adsabs.harvard.edu/abs/2023arXiv230505372C}
}

@ARTICLE{2006ApJ...643..332F,
       author = {{Faucher-Gigu{\`e}re}, Claude-Andr{\'e} and {Kaspi}, Victoria M.},
        title = "{Birth and Evolution of Isolated Radio Pulsars}",
      journal = {\apj},
         year = 2006,
        month = may,
       volume = {643},
       number = {1},
        pages = {332-355},
          doi = {10.1086/501516},
archivePrefix = {arXiv},
       eprint = {astro-ph/0512585},
 primaryClass = {astro-ph},
       adsurl = {https://ui.adsabs.harvard.edu/abs/2006ApJ...643..332F}
}

@ARTICLE{2017arXiv171001191I,
       author = {{IceCube Collaboration} and {Aartsen}, M.~G. and {Ackermann}, M. and {Adams}, J. and {Aguilar}, J.~A. and {Ahlers}, M. and others},
        title = "{The IceCube Neutrino Observatory - Contributions to ICRC 2017 Part II: Properties of the Atmospheric and Astrophysical Neutrino Flux}",
      journal = {arXiv e-prints},
         year = 2017,
        month = oct,
          eid = {arXiv:1710.01191},
        pages = {arXiv:1710.01191},
archivePrefix = {arXiv},
       eprint = {1710.01191},
 primaryClass = {astro-ph.HE},
       adsurl = {https://ui.adsabs.harvard.edu/abs/2017arXiv171001191I}
}

@ARTICLE{2020PhRvD.101j3012C,
       author = {{Capanema}, Antonio and {Esmaili}, Arman and {Murase}, Kohta},
        title = "{New constraints on the origin of medium-energy neutrinos observed by IceCube}",
      journal = {\prd},
         year = 2020,
        month = may,
       volume = {101},
       number = {10},
          eid = {103012},
        pages = {103012},
          doi = {10.1103/PhysRevD.101.103012},
archivePrefix = {arXiv},
       eprint = {2002.07192},
 primaryClass = {hep-ph},
       adsurl = {https://ui.adsabs.harvard.edu/abs/2020PhRvD.101j3012C}
}

@ARTICLE{FermiIGRB,
       author = {{Ackermann}, M. and {Ajello}, M. and {Albert}, A. and {Atwood}, W.~B. and {Baldini}, L. and others},
        title = "{The Spectrum of Isotropic Diffuse Gamma-Ray Emission between 100 MeV and 820 GeV}",
      journal = {\apj},
         year = 2015,
        month = jan,
       volume = {799},
       number = {1},
          eid = {86},
        pages = {86},
          doi = {10.1088/0004-637X/799/1/86},
archivePrefix = {arXiv},
       eprint = {1410.3696},
 primaryClass = {astro-ph.HE},
       adsurl = {https://ui.adsabs.harvard.edu/abs/2015ApJ...799...86A}
}

@ARTICLE{2021JCAP...02..037C,
       author = {{Capanema}, Antonio and {Esmaili}, Arman and {Serpico}, Pasquale Dario},
        title = "{Where do IceCube neutrinos come from? Hints from the diffuse gamma-ray flux}",
      journal = {\jcap},
         year = 2021,
        month = feb,
       volume = {2021},
       number = {2},
          eid = {037},
        pages = {037},
          doi = {10.1088/1475-7516/2021/02/037},
archivePrefix = {arXiv},
       eprint = {2007.07911},
 primaryClass = {hep-ph},
       adsurl = {https://ui.adsabs.harvard.edu/abs/2021JCAP...02..037C}
}

@ARTICLE{1998ApJ...497..759F,
       author = {{Ferri{\`e}re}, Katia},
        title = "{Global Model of the Interstellar Medium in Our Galaxy with New Constraints on the Hot Gas Component}",
      journal = {\apj},
         year = 1998,
        month = apr,
       volume = {497},
       number = {2},
        pages = {759-776},
          doi = {10.1086/305469},
       adsurl = {https://ui.adsabs.harvard.edu/abs/1998ApJ...497..759F}
}

@ARTICLE{2003PASJ...55..191N,
       author = {{Nakanishi}, Hiroyuki and {Sofue}, Yoshiaki},
        title = "{Three-Dimensional Distribution of the ISM in the Milky Way Galaxy: I. The H I Disk}",
      journal = {\pasj},
         year = 2003,
        month = feb,
       volume = {55},
        pages = {191-202},
          doi = {10.1093/pasj/55.1.191},
archivePrefix = {arXiv},
       eprint = {astro-ph/0304338},
 primaryClass = {astro-ph},
       adsurl = {https://ui.adsabs.harvard.edu/abs/2003PASJ...55..191N}
}

@ARTICLE{2006PASJ...58..847N,
       author = {{Nakanishi}, Hiroyuki and {Sofue}, Yoshiaki},
        title = "{Three-Dimensional Distribution of the ISM in the Milky Way Galaxy: II. The Molecular Gas Disk}",
      journal = {\pasj},
         year = 2006,
        month = oct,
       volume = {58},
        pages = {847-860},
          doi = {10.1093/pasj/58.5.847},
archivePrefix = {arXiv},
       eprint = {astro-ph/0610769},
 primaryClass = {astro-ph},
       adsurl = {https://ui.adsabs.harvard.edu/abs/2006PASJ...58..847N}
}

@article{IceCube:2019hmk,
    author = "Aartsen, M. G. and others",
    collaboration = "IceCube",
    title = "{Cosmic ray spectrum and composition from PeV to EeV using 3 years of data from IceTop and IceCube}",
    eprint = "1906.04317",
    archivePrefix = "arXiv",
    primaryClass = "astro-ph.HE",
    doi = "10.1103/PhysRevD.100.082002",
    journal = "Phys. Rev. D",
    volume = "100",
    number = "8",
    pages = "082002",
    year = "2019"
}

@article{ARGO-YBJ:2015cpa,
    author = "Bartoli, B. and others",
    collaboration = "ARGO-YBJ",
    title = "{Study of the Diffuse Gamma-ray Emission From the Galactic Plane With ARGO-YBJ}",
    eprint = "1507.06758",
    archivePrefix = "arXiv",
    primaryClass = "astro-ph.IM",
    doi = "10.1088/0004-637X/806/1/20",
    journal = "Astrophys. J.",
    volume = "806",
    pages = "20",
    month = "6",
    year = "2015"
}

@article{Borione_1998,
	doi = {10.1086/305096},
	url = {https://doi.org/10.1086/305096},
	year = 1998,
	month = {jan},
	publisher = {American Astronomical Society},
	volume = {493},
	number = {1},
	pages = {175--179},
	author = {A. Borione and M. A. Catanese and M. C. Chantell and C. E. Covault and J. W. Cronin and B. E. Fick and L. F. Fortson and J. Fowler and M. A. K. Glasmacher and K. D. Green and D. B. Kieda and J. Matthews and B. J. Newport and D. Nitz and R. A. Ong and S. Oser and D. Sinclair and J. C. van der Velde},
	title = {Constraints on Gamma-Ray Emission from the Galactic Plane at 300 {TeV}},
	journal = {The Astrophysical Journal}
}

@inproceedings{Zhao:2021GJ,
  author = "Zhao, Shiping  and  Zhang, R.  and  Zhang, Y.  and  Yuan, Q.",
  title = "{Measurement of the diffuse gamma-ray emission from Galactic plane with LHAASO-KM2A}",
  doi = "10.22323/1.395.0859",
  booktitle = "Proceedings of 37th International Cosmic Ray Conference {\textemdash} PoS(ICRC2021)",
  year = 2021,
  volume = "395",
  pages = "859"
}

@article{PhysRevLett.115.211101,
  title = {Precision Measurement of the Helium Flux in Primary Cosmic Rays of Rigidities 1.9 GV to 3 TV with the Alpha Magnetic Spectrometer on the International Space Station},
  author = {Aguilar, M. and Aisa, D. and Alpat, B. and Alvino, A. and Ambrosi, G. and others},
  collaboration = {AMS Collaboration},
  journal = {Phys. Rev. Lett.},
  volume = {115},
  issue = {21},
  pages = {211101},
  numpages = {9},
  year = {2015},
  month = {Nov},
  publisher = {American Physical Society},
  doi = {10.1103/PhysRevLett.115.211101},
  url = {https://link.aps.org/doi/10.1103/PhysRevLett.115.211101}
}

@ARTICLE{2023ApJ...949...16S,
       author = {{Schwefer}, Georg and {Mertsch}, Philipp and {Wiebusch}, Christopher},
        title = "{Diffuse Emission of Galactic High-energy Neutrinos from a Global Fit of Cosmic Rays}",
      journal = {\apj},
         year = 2023,
        month = may,
       volume = {949},
       number = {1},
          eid = {16},
        pages = {16},
          doi = {10.3847/1538-4357/acc1e2},
archivePrefix = {arXiv},
       eprint = {2211.15607},
 primaryClass = {astro-ph.HE},
       adsurl = {https://ui.adsabs.harvard.edu/abs/2023ApJ...949...16S}
}

@article{IceCube:2023hou,
    author = "Fuerst, Philipp Michael and others",
    collaboration = "IceCube",
    title = "{Galactic and Extragalactic Analysis of the Astrophysical Muon Neutrino Flux with 12.3 years of IceCube Track Data}",
    doi = "10.22323/1.444.1046",
    journal = "PoS",
    volume = "ICRC2023",
    pages = "1046",
    year = "2023"
}

@inproceedings{Brogi:2021iB,
  author = "Brogi, Paolo  and  Kobayashi, Kazuyoshi",
  title = "{Measurement of the energy spectrum of cosmic-ray helium with CALET on the International Space Station}",
  doi = "10.22323/1.395.0101",
  booktitle = "Proceedings of 37th International Cosmic Ray Conference {\textemdash} PoS(ICRC2021)",
  year = 2021,
  volume = "395",
  pages = "101"
}

@article{Atkin:2017vhi,
    author = "Atkin, E. and others",
    title = "{First results of the cosmic ray NUCLEON experiment}",
    eprint = "1702.02352",
    archivePrefix = "arXiv",
    primaryClass = "astro-ph.HE",
    doi = "10.1088/1475-7516/2017/07/020",
    journal = "JCAP",
    volume = "07",
    pages = "020",
    year = "2017"
}

@article{HAWC:2021bvb,
    author = "Abeysekara, Anushka Udara and others",
    collaboration = "HAWC",
    title = "{Galactic Gamma-Ray Diffuse Emission at TeV energies with HAWC Data}",
    doi = "10.22323/1.395.0835",
    journal = "PoS",
    volume = "ICRC2021",
    pages = "835",
    year = "2021"
}

@article{HESS:2014ree,
    author = "Abramowski, A. and others",
    collaboration = "H.E.S.S.",
    title = "{Diffuse Galactic gamma-ray emission with H.E.S.S}",
    eprint = "1411.7568",
    archivePrefix = "arXiv",
    primaryClass = "astro-ph.HE",
    doi = "10.1103/PhysRevD.90.122007",
    journal = "Phys. Rev. D",
    volume = "90",
    number = "12",
    pages = "122007",
    year = "2014"
}

@article{Abeysekara:2017hyn,
    author = "Abeysekara, A. U. and others",
    title = "{The 2HWC HAWC Observatory Gamma Ray Catalog}",
    eprint = "1702.02992",
    archivePrefix = "arXiv",
    primaryClass = "astro-ph.HE",
    doi = "10.3847/1538-4357/aa7556",
    journal = "Astrophys. J.",
    volume = "843",
    number = "1",
    pages = "40",
    year = "2017"
}

@article{Kheirandish:2019bke,
    author = "Kheirandish, Ali and Wood, Joshua",
    collaboration = "IceCube, HAWC",
    title = "{IceCube Search for Galactic Neutrino Sources based on HAWC Observations of the Galactic Plane}",
    eprint = "1908.08546",
    archivePrefix = "arXiv",
    primaryClass = "astro-ph.HE",
    reportNumber = "PoS-ICRC2019-932",
    doi = "10.22323/1.358.0932",
    journal = "PoS",
    volume = "ICRC2019",
    pages = "932",
    year = "2020"
}

@article{IceCube:2018ndw,
    author = "Aartsen, M. G. and others",
    collaboration = "IceCube",
    title = "{Search for steady point-like sources in the astrophysical muon neutrino flux with 8 years of IceCube data}",
    eprint = "1811.07979",
    archivePrefix = "arXiv",
    primaryClass = "hep-ph",
    doi = "10.1140/epjc/s10052-019-6680-0",
    journal = "Eur. Phys. J. C",
    volume = "79",
    number = "3",
    pages = "234",
    year = "2019"
}

@article{Halzen:2016seh,
    author = "Halzen, Francis and Kheirandish, Ali and Niro, Viviana",
    title = "{Prospects for Detecting Galactic Sources of Cosmic Neutrinos with IceCube: An Update}",
    eprint = "1609.03072",
    archivePrefix = "arXiv",
    primaryClass = "astro-ph.HE",
    doi = "10.1016/j.astropartphys.2016.11.004",
    journal = "Astropart. Phys.",
    volume = "86",
    pages = "46--56",
    year = "2017"
}

@ARTICLE{2022arXiv220315759D,
       author = {{De La Torre Luque}, P. and {Gaggero}, D. and {Grasso}, D. and {Fornieri}, O. and {Egberts}, K. and {Steppa}, C. and {Evoli}, C.},
        title = "{Galactic diffuse gamma rays meet the PeV frontier}",
      journal = {\aap},
         year = 2023,
        month = apr,
       volume = {672},
          eid = {A58},
        pages = {A58},
          doi = {10.1051/0004-6361/202243714},
archivePrefix = {arXiv},
       eprint = {2203.15759},
 primaryClass = {astro-ph.HE},
       adsurl = {https://ui.adsabs.harvard.edu/abs/2023A&A...672A..58D}
}

@ARTICLE{2017JCAP...02..015E,
       author = {{Evoli}, Carmelo and {Gaggero}, Daniele and {Vittino}, Andrea and
         {Di Bernardo}, Giuseppe and {Di Mauro}, Mattia and {Ligorini}, Arianna and
         {Ullio}, Piero and {Grasso}, Dario},
        title = "{Cosmic-ray propagation with DRAGON2: I. numerical solver and astrophysical ingredients}",
      journal = {\jcap},
         year = 2017,
        month = feb,
       volume = {2017},
       number = {2},
          eid = {015},
        pages = {015},
          doi = {10.1088/1475-7516/2017/02/015},
archivePrefix = {arXiv},
       eprint = {1607.07886},
 primaryClass = {astro-ph.HE},
       adsurl = {https://ui.adsabs.harvard.edu/abs/2017JCAP...02..015E}
}

@ARTICLE{2021arXiv211212745P,
       author = {{Porter}, Troy A. and {Johannesson}, Gudlaugur and {Moskalenko}, Igor V.},
        title = "{The GALPROP Cosmic-ray Propagation and Non-thermal Emissions Framework: Release v57}",
      journal = {arXiv e-prints},
         year = 2021,
        month = dec,
          eid = {arXiv:2112.12745},
        pages = {arXiv:2112.12745},
archivePrefix = {arXiv},
       eprint = {2112.12745},
 primaryClass = {astro-ph.HE},
       adsurl = {https://ui.adsabs.harvard.edu/abs/2021arXiv211212745P}
}

@ARTICLE{2013APh....47...54A,
       author = {{Apel}, W.~D. and {Arteaga-Vel{\'a}zquez}, J.~C. and {Bekk}, K. and {Bertaina}, M. and {Bl{\"u}mer}, J. and others},
        title = "{KASCADE-Grande measurements of energy spectra for elemental groups of cosmic rays}",
      journal = {Astroparticle Physics},
         year = 2013,
        month = jul,
       volume = {47},
        pages = {54-66},
          doi = {10.1016/j.astropartphys.2013.06.004},
       adsurl = {https://ui.adsabs.harvard.edu/abs/2013APh....47...54A}
}

@ARTICLE{2019SciA....5.3793A,
       author = {{An}, Q. and {Asfandiyarov}, R. and {Azzarello}, P. and {Bernardini}, P. and {Bi}, X.~J. and others},
        title = "{Measurement of the cosmic ray proton spectrum from 40 GeV to 100 TeV with the DAMPE satellite}",
      journal = {Science Advances},
         year = 2019,
        month = sep,
       volume = {5},
       number = {9},
        pages = {eaax3793},
          doi = {10.1126/sciadv.aax3793},
archivePrefix = {arXiv},
       eprint = {1909.12860},
 primaryClass = {astro-ph.HE},
       adsurl = {https://ui.adsabs.harvard.edu/abs/2019SciA....5.3793A}
}

@ARTICLE{2017ApJ...839....5Y,
       author = {{Yoon}, Y.~S. and {Anderson}, T. and {Barrau}, A. and {Conklin}, N.~B. and {Coutu}, S. and others},
        title = "{Proton and Helium Spectra from the CREAM-III Flight}",
      journal = {\apj},
         year = 2017,
        month = apr,
       volume = {839},
       number = {1},
          eid = {5},
        pages = {5},
          doi = {10.3847/1538-4357/aa68e4},
archivePrefix = {arXiv},
       eprint = {1704.02512},
 primaryClass = {astro-ph.HE},
       adsurl = {https://ui.adsabs.harvard.edu/abs/2017ApJ...839....5Y}
}

@article{Jansson_2012,
	doi = {10.1088/0004-637x/757/1/14},
	url = {https://doi.org/10.1088/0004-637x/757/1/14},
	year = 2012,
	month = {aug},
	publisher = {American Astronomical Society},
	volume = {757},
	number = {1},
	pages = {14},
	author = {Ronnie Jansson and Glennys R. Farrar},
	title = {A {NEW} {MODEL} {OF} {THE} {GALACTIC} {MAGNETIC} {FIELD}},
	journal = {The Astrophysical Journal}
}

@article{Schwefer:2022zly,
    author = "Schwefer, Georg and Mertsch, Philipp and Wiebusch, Christopher",
    title = "{Diffuse Emission of Galactic High-energy Neutrinos from a Global Fit of Cosmic Rays}",
    eprint = "2211.15607",
    archivePrefix = "arXiv",
    primaryClass = "astro-ph.HE",
    reportNumber = "TTK-22-40",
    doi = "10.3847/1538-4357/acc1e2",
    journal = "Astrophys. J.",
    volume = "949",
    number = "1",
    pages = "16",
    year = "2023"
}

@ARTICLE{2017JCAP...06..046M,
       author = {{Merten}, Lukas and {Becker Tjus}, Julia and {Fichtner}, Horst and {Eichmann}, Bj{\"o}rn and {Sigl}, G{\"u}nter},
        title = "{CRPropa 3.1{\textemdash}a low energy extension based on stochastic differential equations}",
      journal = {\jcap},
         year = 2017,
        month = jun,
       volume = {2017},
       number = {6},
          eid = {046},
        pages = {046},
          doi = {10.1088/1475-7516/2017/06/046},
archivePrefix = {arXiv},
       eprint = {1704.07484},
 primaryClass = {astro-ph.IM},
       adsurl = {https://ui.adsabs.harvard.edu/abs/2017JCAP...06..046M}
}

@ARTICLE{2019PhRvL.122r1102A,
       author = {{Calet Collaboration} and {Adriani}, O. and {Akaike}, Y. and {Asano}, K. and {Asaoka}, Y. and {Bagliesi}, M.~G. and others },
        title = "{Direct Measurement of the Cosmic-Ray Proton Spectrum from 50 GeV to 10 TeV with the Calorimetric Electron Telescope on the International Space Station}",
      journal = {\prl},
         year = 2019,
        month = may,
       volume = {122},
       number = {18},
          eid = {181102},
        pages = {181102},
          doi = {10.1103/PhysRevLett.122.181102},
archivePrefix = {arXiv},
       eprint = {1905.04229},
 primaryClass = {astro-ph.HE},
       adsurl = {https://ui.adsabs.harvard.edu/abs/2019PhRvL.122r1102A}
}

@ARTICLE{2012JCAP...01..010B,
       author = {{Blasi}, Pasquale and {Amato}, Elena},
        title = "{Diffusive propagation of cosmic rays from supernova remnants in the Galaxy. I: spectrum and chemical composition}",
      journal = {\jcap},
         year = 2012,
        month = jan,
       volume = {2012},
       number = {1},
          eid = {010},
        pages = {010},
          doi = {10.1088/1475-7516/2012/01/010},
archivePrefix = {arXiv},
       eprint = {1105.4521},
 primaryClass = {astro-ph.HE},
       adsurl = {https://ui.adsabs.harvard.edu/abs/2012JCAP...01..010B}
}

@ARTICLE{2009BRASP..73..564P,
       author = {{Panov}, A.~D. and {Adams}, J.~H. and {Ahn}, H.~S. and {Bashinzhagyan}, G.~L. and {Watts}, J.~W. and others},
        title = "{Energy spectra of abundant nuclei of primary cosmic rays from the data of ATIC-2 experiment: Final results}",
      journal = {Bulletin of the Russian Academy of Sciences, Physics},
         year = 2009,
        month = jun,
       volume = {73},
       number = {5},
        pages = {564-567},
          doi = {10.3103/S1062873809050098},
archivePrefix = {arXiv},
       eprint = {1101.3246},
 primaryClass = {astro-ph.HE},
       adsurl = {https://ui.adsabs.harvard.edu/abs/2009BRASP..73..564P}
}

@article{Ahlers:2013xia,
      author         = "Ahlers, Markus and Murase, Kohta",
      title          = "{Probing the Galactic Origin of the IceCube Excess with
                        Gamma-Rays}",
      journal        = "Phys.Rev.",
      volume         = "D90",
      pages          = "023010",
      doi            = "10.1103/PhysRevD.90.023010",
      year           = "2014",
      eprint         = "1309.4077",
      archivePrefix  = "arXiv",
      primaryClass   = "astro-ph.HE",
      SLACcitation   = "%%CITATION = ARXIV:1309.4077;%%",
}

@article{IceCubeGP,
author = {{IceCube Collaboration}},
title = {Observation of high-energy neutrinos from the Galactic plane},
journal = {Science},
volume = {380},
number = {6652},
pages = {1338-1343},
year = {2023},
doi = {10.1126/science.adc9818},
URL = {https://www.science.org/doi/abs/10.1126/science.adc9818}
}

@ARTICLE{2011Sci...332...69A,
       author = {{Adriani}, O. and {Barbarino}, G.~C. and {Bazilevskaya}, G.~A. and {Bellotti}, R. and {Boezio}, M. and others},
        title = "{PAMELA Measurements of Cosmic-Ray Proton and Helium Spectra}",
      journal = {Science},
         year = 2011,
        month = apr,
       volume = {332},
       number = {6025},
        pages = {69},
          doi = {10.1126/science.1199172},
archivePrefix = {arXiv},
       eprint = {1103.4055},
 primaryClass = {astro-ph.HE},
       adsurl = {https://ui.adsabs.harvard.edu/abs/2011Sci...332...69A}
}

@article{PhysRevLett.114.171103,
  title = {Precision Measurement of the Proton Flux in Primary Cosmic Rays from Rigidity 1 GV to 1.8 TV with the Alpha Magnetic Spectrometer on the International Space Station},
  author = {Aguilar, M. and Aisa, D. and Alpat, B. and Alvino, A. and Ambrosi, G. and others},
  collaboration = {AMS Collaboration},
  journal = {Phys. Rev. Lett.},
  volume = {114},
  issue = {17},
  pages = {171103},
  numpages = {9},
  year = {2015},
  month = {Apr},
  publisher = {American Physical Society},
  doi = {10.1103/PhysRevLett.114.171103},
  url = {https://link.aps.org/doi/10.1103/PhysRevLett.114.171103}
}

@ARTICLE{2006PhRvD..74c4018K,
       author = {{Kelner}, S.~R. and {Aharonian}, F.~A. and {Bugayov}, V.~V.},
        title = "{Energy spectra of gamma rays, electrons, and neutrinos produced at proton-proton interactions in the very high energy regime}",
      journal = {\prd},
         year = 2006,
        month = aug,
       volume = {74},
       number = {3},
          eid = {034018},
        pages = {034018},
          doi = {10.1103/PhysRevD.74.034018},
archivePrefix = {arXiv},
       eprint = {astro-ph/0606058},
 primaryClass = {astro-ph},
       adsurl = {https://ui.adsabs.harvard.edu/abs/2006PhRvD..74c4018K}
}

@ARTICLE{HERMES,
       author = {{Dundovic}, A. and {Evoli}, C. and {Gaggero}, D. and {Grasso}, D.},
        title = "{Simulating the Galactic multi-messenger emissions with HERMES}",
      journal = {\aap},
         year = 2021,
        month = sep,
       volume = {653},
          eid = {A18},
        pages = {A18},
          doi = {10.1051/0004-6361/202140801},
archivePrefix = {arXiv},
       eprint = {2105.13165},
 primaryClass = {astro-ph.HE},
       adsurl = {https://ui.adsabs.harvard.edu/abs/2021A&A...653A..18D}
}

@article{Fermi-LAT:2012edv,
    author = "Ackermann, M. and others",
    collaboration = "Fermi-LAT",
    title = "{Fermi-LAT Observations of the Diffuse Gamma-Ray Emission: Implications for Cosmic Rays and the Interstellar Medium}",
    eprint = "1202.4039",
    archivePrefix = "arXiv",
    primaryClass = "astro-ph.HE",
    doi = "10.1088/0004-637X/750/1/3",
    journal = "Astrophys. J.",
    volume = "750",
    pages = "3",
    year = "2012"
}

@ARTICLE{Tibet21,
       author = { {Tibet AS${\gamma}$ Collaboration} and {Amenomori}, M. and {Bao}, Y.~W. and {Bi}, X.~J. and {Chen}, D. and {Chen}, T.~L. and others},
        title = "{First Detection of sub-PeV Diffuse Gamma Rays from the Galactic Disk: Evidence for Ubiquitous Galactic Cosmic Rays beyond PeV Energies}",
      journal = {\prl},
         year = 2021,
        month = apr,
       volume = {126},
       number = {14},
          eid = {141101},
        pages = {141101},
          doi = {10.1103/PhysRevLett.126.141101},
archivePrefix = {arXiv},
       eprint = {2104.05181},
 primaryClass = {astro-ph.HE},
       adsurl = {https://ui.adsabs.harvard.edu/abs/2021PhRvL.126n1101A}
}

@ARTICLE{2013PhRvD..88l1301M,
       author = {{Murase}, Kohta and {Ahlers}, Markus and {Lacki}, Brian C.},
        title = "{Testing the hadronuclear origin of PeV neutrinos observed with IceCube}",
      journal = {\prd},
         year = 2013,
        month = dec,
       volume = {88},
       number = {12},
          eid = {121301},
        pages = {121301},
          doi = {10.1103/PhysRevD.88.121301},
archivePrefix = {arXiv},
       eprint = {1306.3417},
 primaryClass = {astro-ph.HE},
       adsurl = {https://ui.adsabs.harvard.edu/abs/2013PhRvD..88l1301M}
}

@ARTICLE{2006ApJ...651..142H,
       author = {{Hopkins}, Andrew M. and {Beacom}, John F.},
        title = "{On the Normalization of the Cosmic Star Formation History}",
      journal = {\apj},
         year = 2006,
        month = nov,
       volume = {651},
       number = {1},
        pages = {142-154},
          doi = {10.1086/506610},
archivePrefix = {arXiv},
       eprint = {astro-ph/0601463},
 primaryClass = {astro-ph},
       adsurl = {https://ui.adsabs.harvard.edu/abs/2006ApJ...651..142H}
}

@ARTICLE{2016PhRvL.116g1101M,
       author = {{Murase}, Kohta and {Guetta}, Dafne and {Ahlers}, Markus},
        title = "{Hidden Cosmic-Ray Accelerators as an Origin of TeV-PeV Cosmic Neutrinos}",
      journal = {\prl},
         year = 2016,
        month = feb,
       volume = {116},
       number = {7},
          eid = {071101},
        pages = {071101},
          doi = {10.1103/PhysRevLett.116.071101},
archivePrefix = {arXiv},
       eprint = {1509.00805},
 primaryClass = {astro-ph.HE},
       adsurl = {https://ui.adsabs.harvard.edu/abs/2016PhRvL.116g1101M}
}

@ARTICLE{inelasticity5yr,
       author = {{IceCube Collaboration} and {Aartsen}, M.~G. and {Ackermann}, M. and {Adams}, J. and {Aguilar}, J.~A. and {Ahlers}, M. and others},
        title = "{Measurements using the inelasticity distribution of multi-TeV neutrino interactions in IceCube}",
      journal = {arXiv e-prints},
         year = 2018,
        month = aug,
          eid = {arXiv:1808.07629},
        pages = {arXiv:1808.07629},
archivePrefix = {arXiv},
       eprint = {1808.07629},
 primaryClass = {hep-ex},
       adsurl = {https://ui.adsabs.harvard.edu/abs/2018arXiv180807629I}
}

@ARTICLE{cascade6yr,
       author = {{IceCube Collaboration} and {Aartsen}, M.~G. and {Ackermann}, M. and {Adams}, J. and {Aguilar}, J.~A. and {Ahlers}, M. and others},
        title = "{Characteristics of the diffuse astrophysical electron and tau neutrino flux with six years of IceCube high energy cascade data}",
      journal = {arXiv e-prints},
         year = 2020,
        month = jan,
          eid = {arXiv:2001.09520},
        pages = {arXiv:2001.09520},
archivePrefix = {arXiv},
       eprint = {2001.09520},
 primaryClass = {astro-ph.HE},
       adsurl = {https://ui.adsabs.harvard.edu/abs/2020arXiv200109520I}
}

@ARTICLE{numu95yr,
       author = {{IceCube Collaboration} and {Abbasi}, R. and {Ackermann}, M. and {Adams}, J. and {Aguilar}, J.~A. and {Ahlers}, M. and others},
        title = "{Improved Characterization of the Astrophysical Muon-Neutrino Flux with 9.5 Years of IceCube Data}",
      journal = {arXiv e-prints},
         year = 2021,
        month = nov,
          eid = {arXiv:2111.10299},
        pages = {arXiv:2111.10299},
archivePrefix = {arXiv},
       eprint = {2111.10299},
 primaryClass = {astro-ph.HE},
       adsurl = {https://ui.adsabs.harvard.edu/abs/2021arXiv211110299A}
}

@ARTICLE{HESE75yr,
       author = {{IceCube Collaboration} and {Abbasi}, R. and {Ackermann}, M. and {Adams}, J. and {Aguilar}, J.~A. and {Ahlers}, M. and others},
        title = "{IceCube high-energy starting event sample: Description and flux characterization with 7.5 years of data}",
      journal = {\prd},
         year = 2021,
        month = jul,
       volume = {104},
       number = {2},
          eid = {022002},
        pages = {022002},
          doi = {10.1103/PhysRevD.104.022002},
archivePrefix = {arXiv},
       eprint = {2011.03545},
 primaryClass = {astro-ph.HE},
       adsurl = {https://ui.adsabs.harvard.edu/abs/2021PhRvD.104b2002A}
}

@article{IceCube:2020fpi,
    author = "Abbasi, R. and others",
    collaboration = "IceCube",
    title = "{Detection of astrophysical tau neutrino candidates in IceCube}",
    eprint = "2011.03561",
    archivePrefix = "arXiv",
    primaryClass = "hep-ex",
    doi = "10.1140/epjc/s10052-022-10795-y",
    journal = "Eur. Phys. J. C",
    volume = "82",
    number = "11",
    pages = "1031",
    year = "2022"
}

@article{IceCube:2014stg,
    author = "Aartsen, M. G. and others",
    collaboration = "IceCube",
    title = "{Observation of High-Energy Astrophysical Neutrinos in Three Years of IceCube Data}",
    eprint = "1405.5303",
    archivePrefix = "arXiv",
    primaryClass = "astro-ph.HE",
    doi = "10.1103/PhysRevLett.113.101101",
    journal = "Phys. Rev. Lett.",
    volume = "113",
    pages = "101101",
    year = "2014"
}

@article{IceCube:2016zyt,
    author = "Aartsen, M. G. and others",
    collaboration = "IceCube",
    title = "{The IceCube Neutrino Observatory: Instrumentation and Online Systems}",
    eprint = "1612.05093",
    archivePrefix = "arXiv",
    primaryClass = "astro-ph.IM",
    doi = "10.1088/1748-0221/12/03/P03012",
    journal = "JINST",
    volume = "12",
    number = "03",
    pages = "P03012",
    year = "2017"
}

@article{IceCube-Gen2:2020qha,
    author = "Aartsen, M. G. and others",
    collaboration = "IceCube-Gen2",
    title = "{IceCube-Gen2: the window to the extreme Universe}",
    eprint = "2008.04323",
    archivePrefix = "arXiv",
    primaryClass = "astro-ph.HE",
    doi = "10.1088/1361-6471/abbd48",
    journal = "J. Phys. G",
    volume = "48",
    number = "6",
    pages = "060501",
    year = "2021"
}

@article{IceCube:2016qvd,
    author = "Aartsen, M. G. and others",
    collaboration = "IceCube",
    title = "{The contribution of Fermi-2LAC blazars to the diffuse TeV-PeV neutrino flux}",
    eprint = "1611.03874",
    archivePrefix = "arXiv",
    primaryClass = "astro-ph.HE",
    doi = "10.3847/1538-4357/835/1/45",
    journal = "Astrophys. J.",
    volume = "835",
    number = "1",
    pages = "45",
    year = "2017"
}

@article{IceCube:2023jds,
    author = "Privon, George C. and others",
    collaboration = "IceCube",
    title = "{Search for high-energy neutrino emission from hard X-ray AGN with IceCube}",
    eprint = "2307.15349",
    archivePrefix = "arXiv",
    primaryClass = "astro-ph.HE",
    reportNumber = "PoS-ICRC2023-1032",
    doi = "10.22323/1.444.1032",
    journal = "PoS",
    volume = "ICRC2023",
    pages = "1032",
    year = "2023"
}

@article{IceCube:2019cia,
    author = "Aartsen, M. G. and others",
    collaboration = "IceCube",
    title = "{Time-Integrated Neutrino Source Searches with 10 Years of IceCube Data}",
    eprint = "1910.08488",
    archivePrefix = "arXiv",
    primaryClass = "astro-ph.HE",
    doi = "10.1103/PhysRevLett.124.051103",
    journal = "Phys. Rev. Lett.",
    volume = "124",
    number = "5",
    pages = "051103",
    year = "2020"
}

@article{ANTARES:2022izu,
    author = "Albert, A. and others",
    collaboration = "ANTARES",
    title = "{Hint for a TeV neutrino emission from the Galactic Ridge with ANTARES}",
    eprint = "2212.11876",
    archivePrefix = "arXiv",
    primaryClass = "astro-ph.HE",
    doi = "10.1016/j.physletb.2023.137951",
    journal = "Phys. Lett. B",
    volume = "841",
    pages = "137951",
    year = "2023"
}

@article{IceCube:2022heu,
    author = "Abbasi, R. and others",
    collaboration = "IceCube",
    title = "{Searches for Neutrinos from Large High Altitude Air Shower Observatory Ultra-high-energy \ensuremath{\gamma}-Ray Sources Using the IceCube Neutrino Observatory}",
    eprint = "2211.14184",
    archivePrefix = "arXiv",
    primaryClass = "astro-ph.HE",
    doi = "10.3847/2041-8213/acb933",
    journal = "Astrophys. J. Lett.",
    volume = "945",
    number = "1",
    pages = "L8",
    year = "2023"
}

@article{IceCube:2022jpz,
    author = "Abbasi, R. and others",
    collaboration = "IceCube",
    title = "{Search for High-energy Neutrino Emission from Galactic X-Ray Binaries with IceCube}",
    eprint = "2202.11722",
    archivePrefix = "arXiv",
    primaryClass = "astro-ph.HE",
    doi = "10.3847/2041-8213/ac67d8",
    journal = "Astrophys. J. Lett.",
    volume = "930",
    number = "2",
    pages = "L24",
    year = "2022"
}

@article{IceCube:2020svz,
    author = "Aartsen, M. G. and others",
    collaboration = "IceCube",
    title = "{IceCube Search for High-Energy Neutrino Emission from TeV Pulsar Wind Nebulae}",
    eprint = "2003.12071",
    archivePrefix = "arXiv",
    primaryClass = "astro-ph.HE",
    doi = "10.3847/1538-4357/ab9fa0",
    journal = "Astrophys. J.",
    volume = "898",
    number = "2",
    pages = "117",
    year = "2020"
}

@article{IceCube:2019lzm,
    author = "Aartsen, M. G. and others",
    collaboration = "IceCube",
    title = "{Search for Sources of Astrophysical Neutrinos Using Seven Years of IceCube Cascade Events}",
    eprint = "1907.06714",
    archivePrefix = "arXiv",
    primaryClass = "astro-ph.HE",
    doi = "10.3847/1538-4357/ab4ae2",
    journal = "Astrophys. J.",
    volume = "886",
    pages = "12",
    year = "2019"
}

@article{IceCube:2023ujd,
    author = "Abbasi, R. and others",
    collaboration = "IceCube",
    title = "{Search for Extended Sources of Neutrino Emission in the Galactic Plane with IceCube}",
    eprint = "2307.07576",
    archivePrefix = "arXiv",
    primaryClass = "astro-ph.HE",
    doi = "10.3847/1538-4357/acf713",
    journal = "Astrophys. J.",
    volume = "956",
    number = "1",
    pages = "20",
    year = "2023"
}

@ARTICLE{2023AppSc..13.6433C,
       author = {{Cardillo}, Martina and {Giuliani}, Andrea},
        title = "{The LHAASO PeVatron bright sky: what we learned}",
      journal = {Applied Sciences},
         year = 2023,
        month = may,
       volume = {13},
       number = {11},
          eid = {6433},
        pages = {6433},
          doi = {10.3390/app13116433},
archivePrefix = {arXiv},
       eprint = {2305.10526},
 primaryClass = {astro-ph.HE},
       adsurl = {https://ui.adsabs.harvard.edu/abs/2023AppSc..13.6433C}
}

@ARTICLE{2009ARA&A..47..523H,
       author = {{Hinton}, J.~A. and {Hofmann}, W.},
        title = "{Teraelectronvolt Astronomy}",
      journal = {\araa},
         year = 2009,
        month = sep,
       volume = {47},
       number = {1},
        pages = {523-565},
          doi = {10.1146/annurev-astro-082708-101816},
archivePrefix = {arXiv},
       eprint = {1006.5210},
 primaryClass = {astro-ph.HE},
       adsurl = {https://ui.adsabs.harvard.edu/abs/2009ARA&A..47..523H}
}

@ARTICLE{2012APh....39...61H,
       author = {{Holder}, J.},
        title = "{TeV gamma-ray astronomy: A summary}",
      journal = {Astroparticle Physics},
         year = 2012,
        month = dec,
       volume = {39},
        pages = {61-75},
          doi = {10.1016/j.astropartphys.2012.02.014},
archivePrefix = {arXiv},
       eprint = {1204.1267},
 primaryClass = {astro-ph.HE},
       adsurl = {https://ui.adsabs.harvard.edu/abs/2012APh....39...61H}
}

@ARTICLE{2009PrPNP..63..293B,
       author = {{Bl{\"u}mer}, Johannes and {Engel}, Ralph and {H{\"o}randel}, J{\"o}rg R.},
        title = "{Cosmic rays from the knee to the highest energies}",
      journal = {Progress in Particle and Nuclear Physics},
         year = 2009,
        month = oct,
       volume = {63},
       number = {2},
        pages = {293-338},
          doi = {10.1016/j.ppnp.2009.05.002},
archivePrefix = {arXiv},
       eprint = {0904.0725},
 primaryClass = {astro-ph.HE},
       adsurl = {https://ui.adsabs.harvard.edu/abs/2009PrPNP..63..293B}
}

@article{Lipari:2018gzn,
    author = "Lipari, Paolo and Vernetto, Silvia",
    title = "{Diffuse Galactic gamma ray flux at very high energy}",
    eprint = "1804.10116",
    archivePrefix = "arXiv",
    primaryClass = "astro-ph.HE",
    doi = "10.1103/PhysRevD.98.043003",
    journal = "Phys. Rev. D",
    volume = "98",
    number = "4",
    pages = "043003",
    year = "2018"
}

@ARTICLE{2022A&A...661A..72B,
       author = {{Breuhaus}, M. and {Hinton}, J.~A. and {Joshi}, V. and {Reville}, B. and {Schoorlemmer}, H.},
        title = "{Galactic gamma-ray and neutrino emission from interacting cosmic-ray nuclei}",
      journal = {\aap},
         year = 2022,
        month = may,
       volume = {661},
          eid = {A72},
        pages = {A72},
          doi = {10.1051/0004-6361/202141318},
archivePrefix = {arXiv},
       eprint = {2201.03984},
 primaryClass = {astro-ph.HE},
       adsurl = {https://ui.adsabs.harvard.edu/abs/2022A&A...661A..72B}
}

@ARTICLE{2023A&A...672A..58D,
       author = {{De La Torre Luque}, P. and {Gaggero}, D. and {Grasso}, D. and {Fornieri}, O. and {Egberts}, K. and {Steppa}, C. and {Evoli}, C.},
        title = "{Galactic diffuse gamma rays meet the PeV frontier}",
      journal = {\aap},
         year = 2023,
        month = apr,
       volume = {672},
          eid = {A58},
        pages = {A58},
          doi = {10.1051/0004-6361/202243714},
archivePrefix = {arXiv},
       eprint = {2203.15759},
 primaryClass = {astro-ph.HE},
       adsurl = {https://ui.adsabs.harvard.edu/abs/2023A&A...672A..58D}
}

@ARTICLE{Fang:2023azx,
       author = {{Fang}, Ke and {Gallagher}, John S. and {Halzen}, Francis},
        title = "{The Milky Way revealed to be a neutrino desert by the IceCube Galactic plane observation}",
      journal = {Nature Astronomy},
         year = 2024,
        month = feb,
       volume = {8},
        pages = {241-246},
          doi = {10.1038/s41550-023-02128-0},
archivePrefix = {arXiv},
       eprint = {2306.17275},
 primaryClass = {astro-ph.HE},
       adsurl = {https://ui.adsabs.harvard.edu/abs/2024NatAs...8..241F}
}

@article{IceCube:2023ame,
    author = "Abbasi, R. and others",
    collaboration = "IceCube",
    title = "{Observation of high-energy neutrinos from the Galactic plane}",
    eprint = "2307.04427",
    archivePrefix = "arXiv",
    primaryClass = "astro-ph.HE",
    doi = "10.1126/science.adc9818",
    journal = "Science",
    volume = "380",
    number = "6652",
    pages = "adc9818",
    year = "2023"
}

@article{Murase:2022vqf,
    author = "Murase, Kohta and Mukhopadhyay, Mainak and Kheirandish, Ali and Kimura, Shigeo S. and Fang, Ke",
    title = "{Neutrinos from the Brightest Gamma-Ray Burst?}",
    eprint = "2210.15625",
    archivePrefix = "arXiv",
    primaryClass = "astro-ph.HE",
    doi = "10.3847/2041-8213/aca3ae",
    journal = "Astrophys. J. Lett.",
    volume = "941",
    number = "1",
    pages = "L10",
    year = "2022"
}

@article{LHAASO:2023rpg,
    author = "Cao, Zhen and others",
    collaboration = "LHAASO",
    title = "{The First LHAASO Catalog of Gamma-Ray Sources}",
    eprint = "2305.17030",
    archivePrefix = "arXiv",
    primaryClass = "astro-ph.HE",
    doi = "10.3847/1538-4365/acfd29",
    journal = "Astrophys. J. Suppl.",
    volume = "271",
    number = "1",
    pages = "25",
    year = "2024"
}

@article{IceCube:2022der,
    author = "Abbasi, R. and others",
    collaboration = "IceCube",
    title = "{Evidence for neutrino emission from the nearby active galaxy NGC 1068}",
    eprint = "2211.09972",
    archivePrefix = "arXiv",
    primaryClass = "astro-ph.HE",
    doi = "10.1126/science.abg3395",
    journal = "Science",
    volume = "378",
    number = "6619",
    pages = "538--543",
    year = "2022"
}

@article{Braun:2008bg,
    author = "Braun, Jim and Dumm, Jon and De Palma, Francesco and Finley, Chad and Karle, Albrecht and Montaruli, Teresa",
    title = "{Methods for point source analysis in high energy neutrino telescopes}",
    eprint = "0801.1604",
    archivePrefix = "arXiv",
    primaryClass = "astro-ph",
    doi = "10.1016/j.astropartphys.2008.02.007",
    journal = "Astropart. Phys.",
    volume = "29",
    pages = "299--305",
    year = "2008"
}

@article{IceCube:2018dnn,
      author         = "Aartsen, M. G. and others",
      title          = "{Multimessenger observations of a flaring blazar
                        coincident with high-energy neutrino IceCube-170922A}",
      collaboration  = "IceCube, Fermi-LAT, MAGIC, AGILE, ASAS-SN, HAWC,
                        H.E.S.S., INTEGRAL, Kanata, Kiso, Kapteyn, Liverpool
                        Telescope, Subaru, Swift NuSTAR, VERITAS, VLA/17B-403",
      journal        = "Science",
      volume         = "361",
      year           = "2018",
      number         = "6398",
      pages          = "eaat1378",
      doi            = "10.1126/science.aat1378",
      eprint         = "1807.08816",
      archivePrefix  = "arXiv",
      primaryClass   = "astro-ph.HE",
      SLACcitation   = "%%CITATION = ARXIV:1807.08816;%%"
}

@article{IceCube:2018cha,
    author = "Aartsen, M. G. and others",
    collaboration = "IceCube",
    title = "{Neutrino emission from the direction of the blazar TXS 0506+056 prior to the IceCube-170922A alert}",
    eprint = "1807.08794",
    archivePrefix = "arXiv",
    primaryClass = "astro-ph.HE",
    doi = "10.1126/science.aat2890",
    journal = "Science",
    volume = "361",
    number = "6398",
    pages = "147--151",
    year = "2018"
}

@article{Aartsen:2013jdh,
      author         = "Aartsen, M. G. and others",
      title          = "{Evidence for High-Energy Extraterrestrial Neutrinos at
                        the IceCube Detector}",
      collaboration  = "IceCube",
      journal        = "Science",
      volume         = "342",
      year           = "2013",
      pages          = "1242856",
      doi            = "10.1126/science.1242856",
      eprint         = "1311.5238",
      archivePrefix  = "arXiv",
      primaryClass   = "astro-ph.HE",
      SLACcitation   = "%%CITATION = ARXIV:1311.5238;%%"
}

@article{Fang:2021ylv,
    author = "Fang, Ke and Murase, Kohta",
    title = "{Multimessenger Implications of Sub-PeV Diffuse Galactic Gamma-Ray Emission}",
    eprint = "2104.09491",
    archivePrefix = "arXiv",
    primaryClass = "astro-ph.HE",
    doi = "10.3847/1538-4357/ac11f0",
    journal = "Astrophys. J.",
    volume = "919",
    number = "2",
    pages = "93",
    year = "2021"
}

@ARTICLE{2024NatAs.tmp...54Y,
       author = {{Yan}, Kai and {Liu}, Ruo-Yu and {Zhang}, Rui and {Li}, Chao-Ming and {Yuan}, Qiang and {Wang}, Xiang-Yu},
        title = "{Insights from LHAASO and IceCube into the origin of the Galactic diffuse teraelectronvolt-petaelectronvolt emission}",
      journal = {Nature Astronomy},
         year = 2024,
        month = mar,
          doi = {10.1038/s41550-024-02221-y},
archivePrefix = {arXiv},
       eprint = {2307.12363},
 primaryClass = {astro-ph.HE},
       adsurl = {https://ui.adsabs.harvard.edu/abs/2024NatAs.tmp...54Y}
}

@ARTICLE{2023JCAP...09..027V,
       author = {{Vecchiotti}, V. and {Villante}, F.~L. and {Pagliaroli}, G.},
        title = "{Setting an upper limit for the total TeV neutrino flux from the disk of our Galaxy}",
      journal = {\jcap},
         year = 2023,
        month = sep,
       volume = {2023},
       number = {9},
          eid = {027},
        pages = {027},
          doi = {10.1088/1475-7516/2023/09/027},
archivePrefix = {arXiv},
       eprint = {2306.16305},
 primaryClass = {astro-ph.HE},
       adsurl = {https://ui.adsabs.harvard.edu/abs/2023JCAP...09..027V}
}

@ARTICLE{2023ApJ...956L..44V,
       author = {{Vecchiotti}, V. and {Villante}, F.~L. and {Pagliaroli}, G.},
        title = "{Unveiling the Nature of Galactic TeV Sources with IceCube Results}",
      journal = {\apjl},
         year = 2023,
        month = oct,
       volume = {956},
       number = {2},
          eid = {L44},
        pages = {L44},
          doi = {10.3847/2041-8213/acff60},
archivePrefix = {arXiv},
       eprint = {2307.07451},
 primaryClass = {astro-ph.HE},
       adsurl = {https://ui.adsabs.harvard.edu/abs/2023ApJ...956L..44V}
}

@ARTICLE{2024arXiv240105863R,
       author = {{Roy}, Abhijit and {Joshi}, Jagdish C. and {Cardillo}, Martina and {Sarmah}, Prantik and {Sarkar}, Ritabrata and {Chakraborty}, Sovan},
        title = "{Gamma-rays and Neutrinos from Giant Molecular Cloud Populations in the Galactic Plane}",
      journal = {arXiv e-prints},
         year = 2024,
        month = jan,
          eid = {arXiv:2401.05863},
        pages = {arXiv:2401.05863},
          doi = {10.48550/arXiv.2401.05863},
archivePrefix = {arXiv},
       eprint = {2401.05863},
 primaryClass = {astro-ph.HE},
       adsurl = {https://ui.adsabs.harvard.edu/abs/2024arXiv240105863R}
}

@ARTICLE{2024PhRvD.109d3007A,
       author = {{Ambrosone}, Antonio and {Groth}, Kathrine M{\o}rch and {Peretti}, Enrico and {Ahlers}, Markus},
        title = "{Galactic diffuse neutrino emission from sources beyond the discovery horizon}",
      journal = {\prd},
         year = 2024,
        month = feb,
       volume = {109},
       number = {4},
          eid = {043007},
        pages = {043007},
          doi = {10.1103/PhysRevD.109.043007},
archivePrefix = {arXiv},
       eprint = {2306.17285},
 primaryClass = {astro-ph.HE},
       adsurl = {https://ui.adsabs.harvard.edu/abs/2024PhRvD.109d3007A}
}

@ARTICLE{2024arXiv240305288G,
       author = {{Gagliardini}, Silvia and {Langella}, Aurora and {Guetta}, Dafne and {Capone}, Antonio},
        title = "{Neutrino fluxes from different classes of galactic sources}",
      journal = {arXiv e-prints},
         year = 2024,
        month = mar,
          eid = {arXiv:2403.05288},
        pages = {arXiv:2403.05288},
          doi = {10.48550/arXiv.2403.05288},
archivePrefix = {arXiv},
       eprint = {2403.05288},
 primaryClass = {astro-ph.HE},
       adsurl = {https://ui.adsabs.harvard.edu/abs/2024arXiv240305288G}
}

@article{Fang:2023ffx,
    author = "Fang, Ke and Murase, Kohta",
    title = "{Decomposing the Origin of TeV\textendash{}PeV Emission from the Galactic Plane: Implications of Multimessenger Observations}",
    eprint = "2307.02905",
    archivePrefix = "arXiv",
    primaryClass = "astro-ph.HE",
    doi = "10.3847/2041-8213/ad012f",
    journal = "Astrophys. J. Lett.",
    volume = "957",
    number = "1",
    pages = "L6",
    year = "2023"
}

@article{Strong:1998pw,
    author = "Strong, A. W. and Moskalenko, I. V.",
    title = "{Propagation of cosmic-ray nucleons in the galaxy}",
    eprint = "astro-ph/9807150",
    archivePrefix = "arXiv",
    doi = "10.1086/306470",
    journal = "Astrophys. J.",
    volume = "509",
    pages = "212--228",
    year = "1998"
}

@article{Gaggero:2015xza,
    author = "Gaggero, Daniele and Grasso, Dario and Marinelli, Antonio and Urbano, Alfredo and Valli, Mauro",
    title = "{The gamma-ray and neutrino sky: A consistent picture of Fermi-LAT, Milagro, and IceCube results}",
    eprint = "1504.00227",
    archivePrefix = "arXiv",
    primaryClass = "astro-ph.HE",
    doi = "10.1088/2041-8205/815/2/L25",
    journal = "Astrophys. J. Lett.",
    volume = "815",
    number = "2",
    pages = "L25",
    year = "2015"
}

@article{Aartsen:2020aqd,
    author = "Aartsen, M. G. and others",
    collaboration = "IceCube",
    title = "{Characteristics of the diffuse astrophysical electron and tau neutrino flux with six years of IceCube high energy cascade data}",
    eprint = "2001.09520",
    archivePrefix = "arXiv",
    primaryClass = "astro-ph.HE",
    doi = "10.1103/PhysRevLett.125.121104",
    journal = "Phys. Rev. Lett.",
    volume = "125",
    number = "12",
    pages = "121104",
    year = "2020"
}

@article{Kafexhiu:2014cua,
    author = "Kafexhiu, Ervin and Aharonian, Felix and Taylor, Andrew M. and Vila, Gabriela S.",
    title = "{Parametrization of gamma-ray production cross-sections for pp interactions in a broad proton energy range from the kinematic threshold to PeV energies}",
    eprint = "1406.7369",
    archivePrefix = "arXiv",
    primaryClass = "astro-ph.HE",
    doi = "10.1103/PhysRevD.90.123014",
    journal = "Phys. Rev. D",
    volume = "90",
    number = "12",
    pages = "123014",
    year = "2014"
}

@article{Fang:2014qva,
    author = "Fang, Ke",
    title = "{High-Energy Neutrino Signatures of Newborn Pulsars In the Local Universe}",
    eprint = "1411.2174",
    archivePrefix = "arXiv",
    primaryClass = "astro-ph.HE",
    doi = "10.1088/1475-7516/2015/06/004",
    journal = "JCAP",
    volume = "06",
    pages = "004",
    year = "2015"
}

@ARTICLE{2006ApJ...647..692K,
       author = {{Kamae}, Tuneyoshi and {Karlsson}, Niklas and {Mizuno}, Tsunefumi and {Abe}, Toshinori and {Koi}, Tatsumi},
        title = "{Parameterization of {\ensuremath{\gamma}}, e$^{+/-}$, and Neutrino Spectra Produced by p-p Interaction in Astronomical Environments}",
      journal = {\apj},
         year = 2006,
        month = aug,
       volume = {647},
       number = {1},
        pages = {692-708},
          doi = {10.1086/505189},
archivePrefix = {arXiv},
       eprint = {astro-ph/0605581},
 primaryClass = {astro-ph},
       adsurl = {https://ui.adsabs.harvard.edu/abs/2006ApJ...647..692K}
}

@article{Kachelriess:2015wpa,
    author = "Kachelriess, Michael and Moskalenko, Igor V. and Ostapchenko, Sergey S.",
    title = "{New calculation of antiproton production by cosmic ray protons and nuclei}",
    eprint = "1502.04158",
    archivePrefix = "arXiv",
    primaryClass = "astro-ph.HE",
    doi = "10.1088/0004-637X/803/2/54",
    journal = "Astrophys. J.",
    volume = "803",
    number = "2",
    pages = "54",
    year = "2015"
}

@article{Koldobskiy:2021nld,
    author = "Koldobskiy, S. and Kachelrie\ss{}, M. and Lskavyan, A. and Neronov, A. and Ostapchenko, S. and Semikoz, D. V.",
    title = "{Energy spectra of secondaries in proton-proton interactions}",
    eprint = "2110.00496",
    archivePrefix = "arXiv",
    primaryClass = "astro-ph.HE",
    doi = "10.1103/PhysRevD.104.123027",
    journal = "Phys. Rev. D",
    volume = "104",
    number = "12",
    pages = "123027",
    year = "2021"
}

@article{ARGO-YBJ:2014ikx,
    author = "Bartoli, B. and others",
    collaboration = "ARGO-YBJ",
    title = "{Identification of the TeV Gamma-ray Source ARGO J2031+4157 with the Cygnus Cocoon}",
    eprint = "1406.6436",
    archivePrefix = "arXiv",
    primaryClass = "astro-ph.HE",
    doi = "10.1088/0004-637X/790/2/152",
    journal = "Astrophys. J.",
    volume = "790",
    number = "2",
    pages = "152",
    year = "2014"
}

@article{Abdollahi_2020,
	doi = {10.3847/1538-4365/ab6bcb},
	url = {https://doi.org/10.3847/1538-4365/ab6bcb},
	year = 2020,
	month = {mar},
	publisher = {American Astronomical Society},
	volume = {247},
	number = {1},
	pages = {33},
	author = {S. Abdollahi et al.},
	title = {$Fermi$ Large Area Telescope Fourth Source Catalog},
	journal = {The Astrophysical Journal Supplement Series}
}

@article{IceCube:2021rpz,
    author = "Aartsen, M. G. and others",
    collaboration = "IceCube",
    title = "{Detection of a particle shower at the Glashow resonance with IceCube}",
    doi = "10.1038/s41586-021-03256-1",
    journal = "Nature",
    volume = "591",
    number = "7849",
    pages = "220--224",
    year = "2021"
}

@article{Fang:2022trf,
    author = "Fang, Ke and Gallagher, John S. and Halzen, Francis",
    title = "{The TeV Diffuse Cosmic Neutrino Spectrum and the Nature of Astrophysical Neutrino Sources}",
    eprint = "2205.03740",
    archivePrefix = "arXiv",
    primaryClass = "astro-ph.HE",
    doi = "10.3847/1538-4357/ac7649",
    journal = "Astrophys. J.",
    volume = "933",
    number = "2",
    pages = "190",
    year = "2022"
}

@article{Abdalla:2019dlr,
    author = "Abdalla, H. and others",
    title = "{A very-high-energy component deep in the $\gamma$-ray burst afterglow}",
    eprint = "1911.08961",
    archivePrefix = "arXiv",
    primaryClass = "astro-ph.HE",
    doi = "10.1038/s41586-019-1743-9",
    journal = "Nature",
    volume = "575",
    number = "7783",
    pages = "464--467",
    year = "2019"
}

@article{Waxman:1999ai,
    author = "Waxman, Eli and Bahcall, John N.",
    title = "{Neutrino afterglow from gamma-ray bursts: Similar to 10**18-eV}",
    eprint = "hep-ph/9909286",
    archivePrefix = "arXiv",
    doi = "10.1086/309462",
    journal = "Astrophys. J.",
    volume = "541",
    pages = "707--711",
    year = "2000"
}

@article{Burns:2023oxn,
    author = "Burns, Eric and others",
    title = "{GRB 221009A: The BOAT}",
    eprint = "2302.14037",
    archivePrefix = "arXiv",
    primaryClass = "astro-ph.HE",
    doi = "10.3847/2041-8213/acc39c",
    journal = "Astrophys. J. Lett.",
    volume = "946",
    number = "1",
    pages = "L31",
    year = "2023"
}

\end{document}